\documentclass[tightenlines,showpacs,prd,floatfix,preprintnumbers,amsmath,amssymb,nofootinbib,superscriptaddress]{revtex4}
\usepackage{graphicx}
\usepackage{amsfonts}
\usepackage[normalem]{ulem}
\usepackage{color}
\usepackage{bm}
\usepackage{ulem}
\def\pheq{\phantom{=}}
\newcommand{\Deqn}[1]{{Eq.~(\ref{#1})}}

\newcommand{\beq}{\begin{equation}}
\newcommand{\be}{\begin{equation}}
\newcommand{\eeq}{\end{equation}}
\newcommand{\ee}{\end{equation}}
\newcommand{\bea}{\begin{eqnarray}}
\newcommand{\eea}{\end{eqnarray}}

\newcommand{\beaa}{\begin{eqnarray*}} 
\newcommand{\eeaa}{\end{eqnarray*}}

\newcommand{\dis}{\displaystyle} 
\newcommand{\Lie}{\mbox{\pounds}}
\newcommand{\bsube}{\begin{subequations}}
\newcommand{\esube}{\end{subequations}}

\newcommand{\preto}{\psi_0^{\rm ret}}

\begin{document}

\title{Conservative, gravitational self-force for a particle in circular orbit around a Schwarzschild black hole in a Radiation Gauge}

\author{Abhay G. Shah} 
\email{agshah@uwm.edu}
\affiliation{Center for Gravitation and Cosmology, Department of Physics, 
University of Wisconsin--Milwaukee, P.O. Box 413, Milwaukee, Wisconsin 53201, 
USA}

\author{Tobias S. Keidl} 
\email{tobias.keidl@uwc.edu}
\affiliation{Department of Physics, 
University of Wisconsin--Washington County,USA}

\author{John L. Friedman} 
\email{friedman@uwm.edu}
\affiliation{Center for Gravitation and Cosmology, Department of Physics, 
University of Wisconsin--Milwaukee, P.O. Box 413, Milwaukee, Wisconsin 53201, 
USA}

\author{Dong-Hoon Kim}

\email{ki1313@yahoo.com}
\affiliation{Max-Planck-Institut für Gravitationsphysik, 
Am Mühlenberg 1, D-14476 Golm,
Germany}
\affiliation{Division of Physics, Mathematics, and Astronomy,
California Institute of Technology, Pasadena, CA 91125, USA}
\affiliation{Institute for the Early Universe and Department of Physics, 
Ewha Womans University, Seoul 120-750, South Korea}

\author{Larry R. Price} 
\email{larry@gravity.phys.uwm.edu}
\affiliation{Center for Gravitation and Cosmology, Department of Physics, 
University of Wisconsin--Milwaukee, P.O. Box 413, Milwaukee, Wisconsin 53201, 
USA}

\date{September 21, 2010}

\begin{abstract}
This is the second of two companion papers on computing the self-force in a radiation gauge; more precisely, the method 
uses a radiation gauge for the radiative part of the metric perturbation, together with an arbitrarily chosen gauge for the 
parts of the perturbation associated with changes in black-hole mass and spin and with a shift in the center of mass.
In a test of the method delineated in the first paper, we compute the conservative part of the self-force for a particle in circular 
orbit around a Schwarzschild black hole.  The gauge vector relating our radiation gauge to a Lorenz gauge is helically 
symmetric, implying that the quantity $h_{\alpha\beta}u^\alpha u^\beta $ must have the same value for our radiation gauge 
as for a Lorenz gauge; and we confirm this numerically to one part in $10^{14}$.  As outlined in the first paper, 
the perturbed metric is constructed from a Hertz potential that is in term obtained algebraically from the 
the retarded perturbed spin-2 Weyl scalar, $\psi_0^\textrm{ret}$. We use a mode-sum renormalization and find the 
renormalization coefficients by matching a series in $L=\ell+1/2$ to the large-$L$ behavior of the expression for the self-force 
in terms of the retarded field $h^{\textrm ret}_{\alpha\beta}$; we similarly find the leading renormalization coefficients 
of $h_{\alpha\beta}u^\alpha u^\beta $ and the related change in the angular velocity of the particle due to its self-force.  
We show numerically that the singular part of the self-force has the form $f^{\rm S}_\alpha  =  \langle\nabla_\alpha\rho^{-1}\rangle$, the part of $\nabla_\alpha\rho^{-1}$ that is axisymmetric about 
a radial line through the particle.  This differs only by a constant from its form for a Lorenz gauge.  
It is because we do not use a radiation gauge to describe the change 
in black-hole mass that the singular part of the self-force has no singularity along a radial line through the particle 
and, at least in this example, is spherically symmetric to subleading order in $\rho$.
\end{abstract}

\maketitle

\section{Introduction}

    We present here a first self-force computation in a radiation gauge, following the method 
outlined in a companion paper \cite{sf2} (henceforth Paper I).  The computation has 
been done previously by Barack and Sago \cite{bs07} and by Detweiler \cite{detweiler08}, 
and gauge invariant quantities associated with the conservative part of the self-force 
are compared in their joint paper \cite{sbd08}.  A radiation-gauge approach has the 
advantage that one can use the Teukolsky equation to compute the perturbed metric and the self-force 
for orbits in a Kerr background, and the present paper serves as a test of methods described in 
Paper I for that problem.  

The use of a radiation gauge for the self-force problem 
has been delayed in part because the MiSaTaQuWa renormalization prescription 
\cite{MinoSasaki,quinnwald} was developed for a Lorenz gauge and in part because, 
in a radiation gauge, the linearized metric of a point-particle is singular along 
a ray through the particle.  One can avoid a singularity of this kind by restricting 
the use of a radiation gauge to the part of the perturbation determined 
by the gauge-invariant Weyl scalar $\psi_0$ (or $\psi_4$).  The part of the metric 
perturbation that describes the change in mass and angular momentum of the spacetime
can then be added in any convenient gauge.  The perturbed metric obtained from
from $\psi_0$ is constructed as a sum of angular and time harmonics, defined for 
$r>r_0$ and for $r<r_0$, with $r_0$ the Schwarzschild radial coordinate of the particle. 

Although the $\ell\geq 2$ part of the metric perturbation can be computed more simply 
in a radiation gauge, an analytic computation of the singular field that is to be 
subtracted is significantly more difficult. We avoid the difficulty by replacing the 
analytic computation by a numerical determination of the renormalization coefficients 
that are subtracted in a mode-sum renormalization of the self-force. As a result, 
the efficacy of the method depends on the numerical accuracy with which these coefficients 
can be computed.  We describe the numerical methods used and report tests 
of their accuracy.      

As noted in Paper I, recent work by Gralla \cite{gralla10},
following an earlier derivation of the self-force equations by Gralla and Wald \cite{GrallaWald08},
shows that the first order correction to the geodesic equation, 
\be
 u^\beta\nabla_\beta u^\alpha = a^{\rm ren\, \alpha},
\label{eq:geopert0}
\ee
is obtained from  
\be 
a^{{\rm ret}\,\alpha} = -(g^{\alpha\delta}-u^\alpha u^\delta)
	\left(\nabla_\beta h^{\rm ret}_{\gamma\delta}
	-\frac12\nabla_\delta h^{\rm ret}_{\beta\gamma}\right)u^\beta u^\gamma,
\label{geopert1}\ee
by an angle average in locally inertial coordinates, over a sphere of geodesic 
radius $\rho$ about the particle: 
\be
 a^{\rm ren\, \mu} =  \lim_{\rho\rightarrow 0} \int_{S_\rho} a^{\rm ret\,\mu}\, d\Omega.
\label{eq:aren}\ee
The equation holds in any gauge for which the leading part of $h^{\rm ren}_{\alpha\beta}$ is 
$O(\rho^{-1})$ and has even parity.  Paper I showed that the even-parity condition 
was satisfied in a radiation gauge.  Taking the angle average is equivalent to subtracting 
a field $a^{\rm s}$ (the singular part of the acceleration)
\be
 a^{\rm ren\, \alpha} = a^{\rm ret\, \alpha}-a^{\rm s\, \alpha},
\ee
if $a^{\rm s\, \alpha}$ satisfies the conditions 
(i) The limiting angle average of $a^{\rm s}$, defined as in Eq.~(\ref{eq:aren}), vanishes; 
and (ii) $a^{\rm ret\, \alpha}-a^{\rm s\, \alpha}$  is continuous at the particle. 

The numerical determination of $a^{\rm s\,\alpha}$ from $a^{\rm ret\,\alpha}$ shows, 
with an accuracy close to machine precision, that $a^{\rm s\,\alpha}$ is proportional
to $\langle\nabla^\alpha \rho^{-1}\rangle$ and hence is spherically symmetric.
The result also implies that the singular 
field can be identified with its leading and subleading terms in its mode-sum 
expression as a power series in $L=\ell+1/2$.  It remains an open question whether 
this unexpectedly simple behavior of the singular part of $a^\alpha$ holds 
for generic orbits or for a Kerr background.  

The plan of the paper is as follows.  The perturbed metric is constructed from 
the Weyl scalar $\psi_0^{\rm ret}$, computed as a sum over spin-weighted spherical 
harmonics, and Sec.~\ref{sec2} details the numerical method used to compute the 
retarded radial and angular functions that comprise the sum.  In particular, 
radial integrations using the Teukolsky radial equation and the Sasaki-Nakamura 
form are used and compared.  Sec.~\ref{sec3} describes the computation of 
the retarded metric and the retarded expression for the self force from the  
values of $\psi_0$ for each harmonic.  The conservative part of the self 
force has only a radial component, $f^r$, and it is renormalized by matching 
a power series in $L$ to the sequence of contributions $f^r[\ell]$ from successive
angular harmonics.  We find that the singular field obtained in this way is 
the angular decomposition of $\langle\nabla^\alpha \rho^{-1}\rangle$, with $\rho$ the 
geodesic distance orthogonal to the particle's trajectory.  

 Associated with the perturbed metric of a particle in circular 
orbit is a set of related quantities that are invariant under helically symmetric 
gauge transformations.   The Sago et al. comparison paper \cite{sbd08} 
tabulated values of one of these quantities, and we add the values from 
our radiation-gauge computation to their comparison table.

\section{Computation of $\preto$}
\label{sec2}
\noindent{\em Formalism}

We begin with a brief review of the formalism used in Paper I. We consider a particle of mass $\mathfrak m$ in circular orbit about a Schwarzschild black hole of mass $M$.  We use Schwarzschild coordinates, with the particle at radial coordinate $r_0$ in the $\theta = \pi/2$-plane.
The particle's four-velocity is 
\beq
u^\alpha = u^t(t^\alpha+\Omega\phi^\alpha), 
\eeq
with $t^\alpha$ and $\phi^\alpha$ timelike and rotational Killing vectors, $\Omega= \sqrt{M/r_0^3}$, and 
$u^t = 1/\sqrt{1-3M/r_0}$.  
With a $\delta$-function is normalized by $\int \delta^4(x,z)\sqrt{|g|}d^4x = 1$, its stress-energy tensor is given by  
\bea
T^{\alpha\beta} &=& \frak{m} u^\alpha u^\beta \int \delta^4(x^a-z^a(\tau))d\tau
\nonumber \\
&=& \frak{m}  \frac{u^\alpha u^\beta}{u^t(-g)^{1/2}}\delta(r-r_0)\delta(\cos\theta)\delta(\phi-\Omega t) 
\label{eq:Tab}\eea
where a change of coordinates from $\tau$ to t in the integral for stress-energy tensor is used to 
obtain the second equality. From the addition theorem for spin-weighted spherical harmonics, we have
\beq
T^{\alpha\beta} = \sum_{\ell,m}\frac{\frak{m} u^\alpha u^\beta}{r_0^2 u^t}\delta(r-r_0){\,}_sY_{\ell m}(\theta,\phi){\,}_s\bar Y_{\ell m}(\pi/2,\Omega t).
\eeq  

The Kinnersley tetrad vectors have components 
\beq
(l^\mu) = (1/f(r),1,0,0), \qquad (n^\mu) = \frac{1}{2}(1,-f(r),0,0), \qquad (m^\mu) = \frac{1}{\sqrt{2r}}(0,0,1,i/\sin\theta),
\label{eq:ktet}\eeq
where $f(r):=1-2M/r$; and we denote by $\bf D$, $\bm\Delta$, and $\bm \delta$ the derivative operators along the 
tetrad vectors $l,n$ and $m$, respectively.  The nonvanishing spin-coefficients associated with this tetrad are
\beq
\varrho = -1/r, \qquad \beta = -\alpha = \frac{\cot\theta}{2\sqrt{2}r}, \qquad \mu = -\frac{\Delta}{2r^3}, \qquad \textrm{and}\qquad
	\gamma = \frac{M}{2r^2},
\label{eq:spincoeff}\eeq
 where $\Delta=r^2-2Mr$.
 
The pertubed Weyl scalar $\psi_0= -C_{\alpha\beta\gamma\delta}l^\alpha m^\beta l^\gamma m^\delta$ satisfies the Bardeen-Press equation (the Teukolsky equation for $a=0$), 
\beq
\left[\frac{r^4}\Delta\partial_t^2 -4\left(\frac {Mr^2}\Delta -r\right)\partial_t 
-\Delta^{-2} \frac{\partial}{\partial r} 
\left( \Delta^{3} \frac{\partial} {\partial r} \right) 
- \bar{\eth}\eth \right]\psi_0 = 4\pi r^2 T,  
\label{eq:bp}\ee
where 
\bea
T &:= & -2(\bm \delta-2\beta)\bm \delta T_{\bm{11}} + 4(\bm D - 4\varrho)({\bm\delta}-2\beta)T_{\bm{13}} -2(\bm D-5\varrho)(\bm D-\varrho)T_{\bm{33}} \nonumber\\
&=& T^{(0)} + T^{(1)} + T^{(2)}.
\eea
The superscripts indicate the maximum number of radial derivatives in each term.  Explicitly,   
\bsube\bea
T^{(0)} & =& -\sum_{\ell, m}\frac{{\frak m} u^t}{r_0^4}\delta(r-r_0) [(\ell-1)\ell(\ell+1)(\ell+2)]^{1/2}{}_2Y_{\ell m}(\theta,\phi)\bar{Y}_{\ell m}\left(\frac{\pi}{2},\Omega t\right),\\
T^{(1)} &=& \sum_{\ell, m} \frac{{\frak m}\Omega  u^t}{r_0^2}\left[2i\delta'(r-r_0)  
	+\frac{2m\Omega}{f_0}\delta(r-r_0)\right][(\ell-1)(\ell+2)]^{1/2}
	{}_2Y_{\ell m}(\theta,\phi){}_1\bar{Y}_{\ell m}\left(\frac{\pi}{2},\Omega t\right),\\
T^{(2)} &=& \sum_{\ell m}{\frak m} \Omega^2 u^t\Biggl[\delta''(r-r_0) + \left(\frac{6}{r_0} 
	- \frac{2im\Omega}{f_0}\right)\delta'(r-r_0)\nonumber\\
&\pheq& - \left(\frac{m^2\Omega^2}{f_0^2} +\frac{6im\Omega}{r_0 f_0}+\frac{2imM\Omega}{r_0^2f_0^2}\right)\delta(r-r_0)\Biggr]\,{}_2Y_{\ell m}(\theta,\phi) {}_2\bar{Y}_{\ell m}\left(\frac{\pi}{2},\Omega t\right).
\eea\label{teuksource}\esube

The Weyl scalar has the harmonic decomposition 
\beq
\psi_0 = \sum_{\ell m} e^{-i\omega_m t}R_{0\ell\omega_m}(r) {\,}_2Y_{\ell m}(\theta,\phi),
\qquad \omega_m := m\Omega.   
\eeq  
Its radial functions $R_{0\ell\omega_m}$ and the corresponding radial functions $R_{4\ell\omega_m}$ of $\varrho^{-4}\psi_4$ 
satisfy ($\ell$ and $\omega_m$ subscripts suppressed)
\bea
\Delta R_0^{\prime\prime} + 6(r-M) R_0^\prime + \left[ \frac{\omega^2 r^4}{\Delta} + \frac{4i\omega r^2(r-3M)}{\Delta} - (\ell-2)(\ell+3) \right]R_0 &=& 0,\label{teuk0rad}\\
\Delta R_4^{\prime\prime} - 2(r-M) R_4^\prime + \left[ \frac{\omega^2 r^4}{\Delta} - \frac{4i\omega r^2(r-3M)}{\Delta} - (\ell-1)(\ell+2) \right]R_4 &=& 0,
\label{teuk4rad}
\eea
where the prime denotes a derivative with respect to the radial coordinate $r$. 
Solutions to the above equations are related by
\beq \label{eqnrelation}
R_0 = \frac{\bar{R}_4}{r^4f^2};
\eeq
in this relation, however, $R_0$ and $R_4$ differ by a relative normalization from the radial 
functions of components $\psi_0$ and $\psi_4$ of the same 
vacuum Weyl tensor.  

 We denote by $R_{\rm H}$ and $R_\infty$ solutions to Eq. (\ref{teuk0rad}) that are regular at the horizon and at infinity, respectively, and will write $\dis(^\prime):= \frac d{dr}$. As shown in Sect.~V of paper I,  
the solution to Eq.~(\ref{eq:bp}) is given by  

\bea
\psi_0 &=& \psi_0^{(0)}+\psi_0^{(1)}+\psi_0^{(2)},
\label{eq:psiGT}\eea
where the three terms, corresponding to the three source terms of \Deqn{teuksource}, have the form  
\bsube\bea
\psi_0^{(0)} &=& 4\pi {\frak m} u^t \frac{\Delta_0^2}{r_0^2}\sum_{\ell m}A_{\ell m}[(\ell-1)\ell(\ell+1)(\ell+2)]^{1/2}R_{\rm H}(r_<)R_\infty(r_>){}_2Y_{\ell m}(\theta,\phi)\bar{Y}_{\ell m}\left(\frac{\pi}{2},\Omega t\right), \\
\psi_0^{(1)} &=& 8\pi i{\frak m} \Omega u^t \Delta_0 \sum_{\ell m} A_{\ell m}[(\ell-1)(\ell+2)]^{1/2}
	{}_2Y_{\ell m}(\theta,\phi){}_1\bar{Y}_{\ell m}\left(\frac{\pi}{2},\Omega t\right) \times  \nonumber\\
& & \quad \Bigl\{[im\Omega r_0^2 + 2 r_0]R_{\rm H}(r_<)R_\infty(r_>) 
	+ \Delta_0[R_{\rm H}'(r_0)R_\infty(r)\theta(r-r_0) 	 + R_{\rm H}(r)R_\infty'(r_0)\theta(r_0-r)]\Bigr\},\\
\psi_0^{(2)} &=& -4\pi {\frak m}\Omega^2 u^t \sum_{\ell m} A_{\ell m}
	{}_2Y_{\ell m}(\theta,\phi){}_2\bar{Y}_{\ell m}\left(\frac{\pi}{2},\Omega t\right) \times 
\nonumber\\ 
& & \biggl\{[30r_0^4 - 80Mr_0^3 + 48M^2r_0^2 - m^2\Omega^2 r_0^6 -2\Delta_0^2 - 24\Delta_0 r_0(r_0-M)+ 6im\Omega r_0^4(r_0-M)]
	R_{\rm H}(r_<)R_\infty(r_>)
 \nonumber\\ 
& & \qquad  + 2(6r_0^5 - 20Mr_0^4 + 16M^2r_0^3 - 3r_0\Delta_0^2 + im\Omega \Delta_0 r_0^4)[R_{\rm H}'(r_0)R_\infty(r)\theta(r-r_0) 
 + R_\infty'(r_0)R_{\rm H}(r)\theta(r_0-r)] \nonumber\\ 
& & \qquad + r_0^2\Delta_0^2[R_{\rm H}''(r_0)R_\infty(r)\theta(r-r_0) + R_\infty''(r_0)R_{\rm H}(r)\theta(r_0-r)+\textrm{W}[R_{\rm H}(r),R_\infty(r)]\delta(r-r_0)]\Biggr\}.
\eea\esube
Here the Wronskian-related quantity, 
\beq
A_{\ell m} := \frac{1}{\Delta^3 (R_{\rm H} R_\infty^\prime - R_\infty R_{\rm H}^\prime)},
\eeq
is a constant, independent of $r$.\\

\noindent{\em Numerical method}\\

To compute $\preto$, we use a 7th order Runge-Kutta routine to integrate the radial Teukolsky equation,  matching the radial function to a power series expansion near the horizon and infinity. Because our renormalization method relies on numerical extraction of 
the renormalization coefficients, we work to high numerical precision and check the results by 
comparing independent codes based on the Teukolsky form of the radial equation (\ref{teuk0rad}) and on 
the Sasaki-Nakamura form.  We find the spin-weighted spherical harmonics ${\,}_sY_{\ell m}(\frac{\pi}{2},0)$ to similarly high precision.   

Because one of solutions for $R_0$ diverges at the horizon, we integrate $R_4$ from the horizon and $R_0$ from infinity. We match near the horizon and at large $r$, respectively, to the series expansions
\bea \label{bound_cond_of_T}
R_4^\textrm{H} &=& e^{i\omega r^*}\sum_n a_n \left(\frac{r}{M}-2\right)^{n}, \\
R_{0\infty} &=&  e^{i\omega r^*}\sum_n \frac{b_n}{(r/M)^{n+5}}, 
\eea
where $r^* = r+2M \log(r/2M-1)$. The expansion coefficients are found from a 3-term recurrence relation,
\bea
a_n &=& \frac{-2i\omega M(n-5)a_{n-2}+(\ell^2+\ell-6+5n-n^2-8in\omega M +20i\omega M)a_{n-1}}{2n(n-2+4i\omega M)}, \\
b_n &=& \frac{i}{2n\omega M}\left( (6+8n+2n^2)b_{n-2} + (\ell^2+\ell-2-3n-n^2)b_{n-1} \right),
\eea
obtained by substitution in Eqs.~(\ref{teuk0rad}) and (\ref{teuk4rad}).
The integrations from the horizon and from infinity yield the two independent solutions to the homogenous Teukolsky equation that we have labeled $R_{\rm H}$ and $R_\infty$. By using the outgoing wave solution at the event horizon as an initial condition for $R_4^H$, we get the correct boundary condition (ingoing at horizon) for s = +2 radial solution on using Eq.~(\ref{eqnrelation}) i.e. 
\beq
R_0^H \sim \frac{e^{-i\omega r*}}{r^4 f^2}.
\eeq
\\

The numeric results are compared with the numerical solution obtained from
Sasaki-Nakamura equation for consistency.  In this case, we integrate
\beq\label{theaboveeqn}
\frac{d^2X_{\ell,m,\omega}}{dr^{*2}} = U(r) X_{\ell,m,\omega},
\eeq
where 
\beq
U(r) = \frac{12M^2-2Mr(\ell^2+\ell+3)+r^2(\ell^2+\ell-r^2\omega^2)}{r^4}.
\eeq
The function $X_{\ell,m,\omega}$ is related to the homogenous solution of the $s = -2$ radial, Teukolsky equation by
\beq
 _{4}R_{\ell,m,\omega} = \frac{2\Delta(r-3M+ir^2\omega)}{\eta r}X^\prime_{\ell,m,\omega} + \left(\frac{l(l+1)\Delta}{\eta r} - \frac{6M\Delta}{\eta r^2} - \frac{2r\omega(3iM-ir+r^2\omega)}{\eta} \right) X_{\ell,m,\omega},
\eeq
where $\eta = (l-1)l(l+1)(l+2)-12iM\omega $.
We use a 7th order Runge-Kutta routine to integrate Eq.(\ref{theaboveeqn}) with the boundary conditions
\bea \label{bound_cond_of_SN}
X^H_{\ell,m,\omega} & = & e^{i\omega r*} \sum_n c_n \left(\frac r{M}-2\right)^n, \\
X^{\infty}_{\ell,m,\omega} & = & e^{-i\omega r*} \sum_n \frac{d_n}{(r/M)^n},
\eea
where $c_n = d_n$ = 0 for $n < 0$.
The values of $c_n$ and $d_n$ are calculated from the following recurrence relations:  
\bea
c_n &=& -\frac{i(n-3)M\omega}{2n(n+4iM\omega)}c_{n-3} + \frac{\ell^2 
	+ \ell -(n-2)(n-3+12iM\omega)}{4n(n+4iM\omega)}c_{n-2},  \\ \nonumber
 && +\frac{\ell^2 +\ell -2n^2+5n-6-12i(n-1)M\omega}{2n(n+4iM\omega)}c_{n-1},\\
d_n &=&  -i\frac{(n-3)(n+1)}{2nM\omega}d_{n-2} -i\frac{(\ell+n)(\ell-n+1)}{2nM\omega}d_{n-1}.
\eea

The outgoing (ingoing) boundary condition at infinity (event horizon) is used in combination with Eq.~(\ref{eqnrelation}) which relates the $s=2$ and $s=-2$ radial solutions.\\

The accuracy of the homogenous solutions is monitored by the constancy of $A_{\ell m}^{-1} = \Delta^3\ W( R_{\rm H},R_\infty)$. Table \ref{table1} shows the fractional standard deviation (f.s.d.), or 
the departure of $\Delta^3 W$ from its average value. 
A comparison of the $s=\pm2$ radial functions given by the Teukolsky equation and the Sasaki-Nakamura equation (Eqs.~(\ref{teuk0rad}) and (\ref{teuk4rad})) are included.  Measured in this way, 
direct integration of the Teukolsky form yields 13-digit accuracy.  
Table II reports a second check of numerical accuracy by directly comparing the radial functions computed from the two forms of the radial equation.     

\begin{center}
\begin{table}
  \begin{tabular}{|c|c|c|c|}
    \hline
                 \multicolumn{1}{|p{3.0cm}|}
                 {\centering $\ell$}
                & \multicolumn{1}{|p{3.0cm}|}
                 {\centering $M\omega$}
                & \multicolumn{1}{|p{3.0cm}|}
                 {\centering f.s.d.(T-eqn)} 
                 & \multicolumn{1}{|p{3.0cm}|}
                 {\centering f.s.d.(SN-eqn)} \\

    \hline
       3	&	0.001   	&	$4.2\times10^{-14}$  &  $2.8\times10^{-14}$\\
       2    &   0.136083	&	$2.8\times10^{-14}$  &  $9.7\times10^{-13}$\\
       5    &   0.000353553	&	$4.4\times10^{-14}$  &  $2.7\times10^{-14}$\\
       5	&	0.340207	&	$3.5\times10^{-14}$  &  $1.3\times10^{-12}$\\
       10	&	0.00108866	&	$5.6\times10^{-14}$  &  $4.3\times10^{-14}$\\
       9	&	0.612372	&	$4.3\times10^{-14}$  &  $1.0\times10^{-12}$\\
       14	&	0.000715542	&	$6.2\times10^{-14}$  &  $5.5\times10^{-14}$\\
       15	&	1.02062		&	$5.1\times10^{-14}$  &  $1.1\times10^{-12}$\\
       20	&	0.000929429	&	$7.8\times10^{-14}$  &  $7.4\times10^{-14}$\\
       21	&	1.42887		&	$5.9\times10^{-14}$  &  $1.4\times10^{-12}$\\
       25	&	0.000707107	&	$9.2\times10^{-14}$  &  $9.7\times10^{-14}$\\
       25	&	1.70103		&	$7.2\times10^{-14}$  &  $1.1\times10^{-12}$\\
       30	&	0.002		&	$1.0\times10^{-13}$  &  $1.2\times10^{-13}$\\
       31	&	2.10928		&	$8.0\times10^{-14}$  &  $1.6\times10^{-12}$\\
       35	&	0.000544331	&	$1.2\times10^{-13}$  &  $1.2\times10^{-13}$\\
       34	&	2.31341		&	$8.9\times10^{-14}$  &  $1.1\times10^{-12}$\\
       40	&	0.000988212	&	$1.3\times10^{-13}$  &  $1.5\times10^{-13}$\\
       40	&	2.72166		&	$9.6\times10^{-14}$  &  $1.6\times10^{-12}$\\
       45	&	0.000603682	&	$1.5\times10^{-13}$  &  $1.7\times10^{-13}$\\
       46	&	3.1299		&	$1.1\times10^{-13}$  &  $1.3\times10^{-12}$\\
       50	&	0.000353553	&	$1.6\times10^{-13}$  &  $1.8\times10^{-13}$\\
       50	&	3.40207		&	$1.2\times10^{-13}$  &  $1.5\times10^{-12}$\\
       55	&	0.00067466	&	$1.8\times10^{-13}$  &  $2.0\times10^{-13}$\\
       55	&	3.74228		&	$1.3\times10^{-13}$  &  $1.1\times10^{-12}$\\
       60	&	0.001		&	$2.0\times10^{-13}$  &  $2.3\times10^{-13}$\\
       61	&	4.15052		&	$1.3\times10^{-13}$  &  $1.5\times10^{-12}$\\
       65	&	0.000544331	&	$2.2\times10^{-13}$  &  $2.5\times10^{-13}$\\
       64	&	4.35465		&	$1.4\times10^{-13}$  &  $1.5\times10^{-12}$\\
       71	&	0.000494106	&	$2.3\times10^{-13}$  &  $2.8\times10^{-13}$\\
       70	&	4.7629		&	$1.7\times10^{-13}$  &  $1.4\times10^{-12}$\\
       75	&	0.000637528	&	$2.5\times10^{-13}$  &  $3.1\times10^{-13}$\\
       74	&	5.03506		&	$1.8\times10^{-13}$  &  $1.5\times10^{-12}$\\
       80	&	0.000471818	&	$2.8\times10^{-13}$  &  $3.5\times10^{-13}$\\
       80	&	5.44331		&	$1.9\times10^{-13}$  &  $1.5\times10^{-12}$\\
       85	&	0.000707107	&	$3.0\times10^{-13}$  &  $3.9\times10^{-13}$\\
       85	&	5.78352		&	$2.1\times10^{-13}$  &  $1.0\times10^{-12}$\\

    \hline
  \end{tabular}
  \caption{ Accuracy of radial integration measured by fractional standard deviation (f.s.d) of 
$\Delta^3 W$ from its average value, for specified $\ell$ and $\omega$. The third and fourth columns list the f.s.d. obtained by integrating the Teukolsky and Sasaki-Nakamura forms of the radial equation, respectively. The frequencies are chosen to lie between the maximum and minimum of $m\Omega$, between the value 
$M\omega = \ell/6^{3/2}$ at the ISCO and $M\omega \sim 1/150^{3/2}$, 
at $r_0\sim 150 M$, $m=1$. 
The f.s.d. in the region r = 6M to 20M is usually twice the average shown above.}
  \label{table1}\end{table}
\end{center}

\begin{center}
\begin{table}
  \begin{tabular}{|c|c|c|c|}
    \hline
                 \multicolumn{1}{|p{3.0cm}|}
                 {\centering $r_0/M$}
                & \multicolumn{1}{|p{3.0cm}|}
                 {\centering $\ell$}
                & \multicolumn{1}{|p{3.0cm}|}
                 {\centering $m$} 
                 & \multicolumn{1}{|p{3.0cm}|}
                 {\centering fractional difference} \\

    \hline
       100	&	2	&	1	&	5.9$\times10^{-13}$\\
       10	&	2	&	2	&	4.6$\times10^{-15}$\\
       6	&	2	&	1	&	1.9$\times10^{-13}$\\
       80	&	75	&	75	&	1.7$\times10^{-15}$\\
       80	&	20	&	1	&	3.2$\times10^{-15}$\\
       13	&	5	&	4	&	1.6$\times10^{-14}$\\
       10	&	3	&	1	&	2.3$\times10^{-14}$\\
       8	&	15	&	14	&	5.0$\times10^{-15}$\\
       8	&	12	&	6	&	1.6$\times10^{-14}$\\
       70	&	3	&	2	&	1.8$\times10^{-15}$\\
       6	&	4	&	1	&	4.1$\times10^{-13}$\\
       50	&	4	&	1	&	1.0$\times10^{-13}$\\
       6	&	10	&	9	&	4.7$\times10^{-13}$\\
       70	&	9	&	8	&	1.7$\times10^{-15}$\\
       6	&	25	&	1	&	4.0$\times10^{-16}$\\
       7	&	25	&	25	&	2.4$\times10^{-15}$\\
       75	&	19	&	18	&	5.0$\times10^{-15}$\\
       6	&	20	&	20	&	2.0$\times10^{-14}$\\
       72	&	85	&	78	&	1.5$\times10^{-15}$\\
       6	&	85	&	85	&	2.4$\times10^{-14}$\\
       7	&	85	&	5	&	4.2$\times10^{-14}$\\
       60	&	75	&	1	&	$4.0\times10^{-16}$\\
       10	&	75	&	71	&	$2.2\times10^{-15}$\\
       15	&	50	&	38	&	$1.4\times10^{-15}$\\
       6	&	40	&	25	&	$7.2\times10^{-15}$\\
       6	&	25	&	15	&	$7.1\times10^{-15}$\\
       30	&	25	&	1	&	$6.0\times10^{-16}$\\
       20	&	25	&	25	&	$7.8\times10^{-15}$\\
       10	&	2	&	1	&	$3.9\times10^{-14}$\\

    \hline
  \end{tabular}
\caption{For each listed value of $\ell$, $m$ and $r_0$, we give the fractional difference 
between the radial functions at $r_0$, obtained by integrating the Teukolsky 
and the Sasaki-Nakamura forms of the radial equation, with frequency 
$\omega = m\Omega = m M^{1/2} r_0^{-3/2}$.}
  \end{table}
\end{center}

To calculate ${\,}_sY_{\ell m}$ to high precision, we used the following analytical forms of 
spin-weighted harmonics at $\theta=\pi/2$.  Introducing the symbol  
$\displaystyle 
 	e_{\ell,m} := \begin{cases} 1 & \ell+m\ \mbox{ even}\\
		0, & \ell+m\ \mbox{ odd}\ ,
\end{cases}
$
we can write
\bea
Y_{\ell m}(\frac{\pi}{2},0) &=& (-1)^{(\ell + m)/2}\frac{\sqrt{(2\ell +1)(\ell -m)!(\ell +m)!}}{\sqrt{4\pi}(\ell -m)!!(\ell +m)!!} e_{\ell,m} ,
\\  
\nonumber\\
{\,}_1Y_{\ell m}(\frac{\pi}{2},0) 
&=& (-1)^{(\ell + m)/2} \sqrt{\frac{(2\ell +1)(\ell -m)!(\ell +m)!}{4\pi \ell (\ell +1)}} 
\left[\frac{m\,e_{\ell,m}}{(\ell -m)!!(\ell+m)!!}-\frac{i\,e_{\ell,m+1}}{(\ell-m-1)!!(\ell+m-1)!!} \right] ,
\\ 
{\,}_2Y_{\ell m}(\frac{\pi}{2},0) 
&=& (-1)^{(\ell + m)/2} \sqrt{\frac{(2\ell +1)(\ell -m)!(\ell +m)!}{4\pi (\ell-1)\ell (\ell +1)(\ell+2)}}
\left[ \frac{\left[ 2m^2-\ell(\ell+1)\right]e_{\ell,m}}{(\ell-m)!!(\ell+m)!!} - \frac{2i\,m\,e_{\ell,m+1}}{(\ell-m-1)!!(\ell+m-1)!!} \right] .
\eea

 In this way, we obtain values of $\psi_0^\textrm{ret}$ with accuracy of 1 part in $10^{13}$.

\section{Metric pertubation and self-force}
\label{sec3}

We use the outgoing radiation gauge (ORG) satisfying the conditions
\beq
h_{\alpha\beta}n^\alpha=0, \quad h = 0. 
\eeq
We find the retarded perturbed metric from a Hertz potential $\Psi^\textrm{ret}$ satisfying
\beq
8 \preto = \eth^4 \bar\Psi^{\rm ret} + 12M\partial_t \Psi^{\rm ret}, 
\label{eq:Psi}\eeq
whose algebraic solution for each angular harmonic is given by 
\beq
\Psi_{\ell m} = 8 \frac{(-1)^m (\ell+2)(\ell+1)\ell(\ell-1)\bar\psi_{\ell,-m}
	+ 12 i m M \Omega \psi_{\ell m} }{ [(\ell+2)(\ell+1)\ell(\ell-1)]^2 + 144 m^2 M^2 \Omega ^2}
\eeq
where $\Psi = \sum_{\ell,m}\Psi_{\ell m}(r){\,}_2Y_{\ell m}(\theta,\phi)e^{-im\Omega t}$ and $\psi_0 = \sum_{\ell,m}\psi_{\ell m}(r){\,}_2Y_{\ell m}(\theta,\phi)e^{-im\Omega t}$.

The ORG form of the metric pertubation $h_{\alpha\beta}$ is
\bea
h_{\alpha\beta}&=& r^4\{\, n_\alpha n_\beta (\bar{\bm\delta}+2\beta)(\bar{\bm\delta}+4\beta)
		+\bar{m}_\alpha\bar{m}_\beta(\bm\Delta+5\mu-2\gamma)(\bm\Delta+\mu-4\gamma) \nonumber\\
&\pheq& - n_{(\alpha}\bar{m}_{\beta)} \left[(\bar{\bm \delta} +4\beta) (\bm \Delta + \mu - 4\gamma)+
(\bm \Delta + 4\mu -4\gamma)(\bar{\bm \delta} +4\beta) \right]\, \} \Psi+\rm{c.c.}\,,
\label{mpschw}
\eea
where the spin coefficients are given by Eq.~(\ref{eq:spincoeff}).
After writing $\bm\delta$ in terms of $\eth$, the tetrad components take the form
\bea
h_{\bf 11} &=& \frac{r^2}{2}(\bar{\eth}^2\Psi + \eth^2\overline\Psi), \\ 
h_{\bf 33} &=& r^4\left[\frac{\partial_t^2 -2f\partial_t\partial_r+f^2\partial_r^2}{4} - \frac{3(r-M)}{2r^2}\partial_t
+ \frac{f(3r-2M)}{2r^2}\partial_r + \frac{r^2-2M^2}{r^4}\right]\Psi, \\
h_{\bf 13} &=& -\frac{r^3}{2\sqrt{2}}\left(\partial_t-f\partial_r-\frac{2}{r}\right)\bar{\eth}{\Psi}.
\eea

Finally, the self-acceleration, 
\be
a^\alpha = -(g^{\alpha\delta} - u^\alpha u^\delta)
	\left(\nabla_\beta h^{\rm ren}_{\gamma\delta}
	-\frac12\nabla_\delta h^{\rm ren}_{\beta\gamma}\right)u^\beta u^\gamma,
\label{geopert}\ee
written in terms of these tetrad components, is given by
\bea
a^r&=& (u^t)^2\left\{f_0^2\left[ \frac1{16}f_0 \bm D +\frac38\,\bm\Delta  
	+\frac i8\Omega(\eth-\bar\eth)-\frac12\frac{M }{r_0^2} \right] h_{\bf11}\right.
 \nonumber\\
&&\left.\phantom{xxxxx}+f_0\left[\left(\frac18\frac{M}{r_0} \bm D - \frac14\frac{M}{r_0f_0}\bm\Delta 
	+ \frac12\frac{M}{r_0^2} \right)h_{\bf33} 
 + \left(-\frac i{\sqrt{2}}\Omega r_0 \bm\Delta + \frac1{4\sqrt2}\frac{M}{r_0^2}(\eth-\bar{\eth})
	+\frac i{2\sqrt2}\Omega \right)h_{\bf13}+c.c.\right]\right\}. 
\label{eq:ar}\eea
The harmonic decomposition, $\dis\Psi = \sum_{\ell m} \Psi_{\ell m}(r) {\,}_2Y_{\ell m}(\theta,\phi) e^{-im\Omega t}$, of the Hertz potential gives a corresponding decomposition 
of $a^r$.  The contribution to $a^r$ from the harmonic $\Psi_{\ell m}$, however, is not a single angular 
harmonic because the particle's velocity $u^\alpha$ involves the Killing vector $\phi^\alpha$, 
an axial $\ell = 1$ vector field on the 2-sphere; and $u^\alpha$ occurs quadratically 
in Eq.~(\ref{geopert}) for $a^r$.   Instead, the terms in Eq.~(\ref{eq:ar}) include terms from
$h_{\bf11}$ and $\bar{\eth}h_{\bf 13}$ proportional to $Y_{\ell m}$; terms from $\eth h_{\bf 11}$ and $h_{\bf 13}$ proportional 
to $_1\!Y_{\ell m}$; and terms from $\eth h_{\bf13}$ and $h_{\bf 33}$ proportional to $_2\!Y_{\ell m}$.
A virtue of our numerical renormalization procedure is that we need not rewrite these 
latter terms as sums of spin-weight zero harmonics.

We organize the computation by writing the right side of Eq.~(\ref{eq:ar}) as a sum of six terms for which 
the magnitude $|s|$ of the spin-weight increases from one red square bracket to the next:   
\bea
a^r&=& \left[ (u^t)^2f_0^2\left( \frac1{16}f_0 \bm D +\frac38\,\bm\Delta -\frac12\frac{M }{r_0^2} \right) h_{\bf11} 
\ -\ \frac1{4\sqrt2}(u^t)^2\frac{Mf_0}{r_0^2}(\bar\eth h_{\bf 13} + \eth h_{\bf 14} ) \right]
 \nonumber\\ 
&&+\left[\frac i8(u^t)^2f_0^2\Omega(\eth-\bar\eth)h_{\bf11}\ -\ 
\frac i{\sqrt2}(u^t)^2\Omega f_0(r_0\bm\Delta -\frac12)(h_{\bf 13}-h_{\bf 14})\right]
\nonumber\\
&&  +\left[\frac1{4\sqrt2}(u^t)^2\frac{f_0 M}{r_0^2}(\eth h_{\bf13}+\bar\eth h_{\bf 14})
+(u^t)^2\frac{Mf_0}{r_0}\left(\frac18 \bm D - \frac1{4f_0}\bm\Delta 
	+ \frac1{2r_0} \right)(h_{\bf33}+h_{\bf44})\right]
\label{eq:ar1}\\
&=& \sum_{i=1}^6 a^r_i.	
\eea
The subscript $i$ ($i=1,\ldots, 6$) 
in the symbol $a^r_i$ refers to the location of the term on the right side of Eq (\ref{eq:ar1}).
We denote by $a^r[\ell]$ the contribution to $a^r$ from the restriction of $\Psi$ to the 
$\ell$-subspace, writing
\beq
a^r[\ell]= \sum_{i=1}^{6}a^r_i[\ell]. 
\label{eq:ar-l}\eeq

The harmonics $a^r_i[\ell]$ have finite limits as $r\rightarrow r_0^\pm$. We evaluate $a^r$ at a 
particle position $P$ with coordinates $t=0$, $\theta=\pi/2$, and $\phi=0$, where the value of $_s\!Y_{\ell m}$ is real.  
The harmonics then have at $P$ (for either choice of limit) the forms 
\bsube\bea
a^r_1[\ell] &=& - \frac18[(\ell-1)\ell(\ell+1)(\ell+2)]^{1/2}(u^t)^2f_0^2r_0^2
	\left[-2\partial_t+f_0\left(\partial_r+\frac2r\right)+4\frac M{r_0^2}\right]\sum_{m} \textrm{Re}\,(\Psi_{\ell m}) Y_{\ell m},
\\
a^r_2[\ell]&=& \frac18[(\ell-1)\ell(\ell+1)(\ell+2)]^{1/2}(u^t)^2Mr_0f_0
  \sum_m\left[m\Omega\,\textrm{Im}(\Psi_{\ell m})-\left(f_0\partial_r+\frac2{r_0}\right)\textrm{Re}(\Psi_{\ell m})\right]Y_{\ell m},
\\
a_3^r[\ell]&=& - \frac18\ell(\ell+1)[(\ell-1)(\ell+2)]^{1/2}(u^t)^2f_0^2r_0^2\Omega 
		\sum_{m}\textrm{Im}(\Psi_{\ell m})(_1\!Y_{\ell m}+{}_{-1}\!Y_{\ell m}),
\\
a^r_4[\ell] &=& \frac14[(\ell-1)(\ell+2)]^{1/2}(u^t)^2f_0r_0^4\Omega
	\left(\partial_t^2-2f_0\partial_t\partial_r+f_0^2\partial_r^2  - \frac{3}{r_0}(1+f_0)\partial_t + \frac{2f_0}{r_0}\left(3-2\frac{M}{r_0}\right)\partial_r \right. \\ \nonumber
	&&\phantom{xxxxx} \left. + \frac{2}{r_0^2}(1+2f_0)\right) \sum_m \, \textrm{Im}\,\Psi_{\ell m}\,{}_1\!Y_{\ell m},
\\
a^r_5[\ell] &=&  \frac18(\ell-1)(\ell+2)(u^t)^2f_0Mr_0\sum_m(m\Omega\textrm{Im}(\Psi_{\ell m})-f_0\partial_r\textrm{Re}(\Psi_{\ell m})-2/r\textrm{Re}(\Psi_{\ell m}))
		 \,  {}_2\!Y_{\ell m},
\\
a^r_6[\ell] &=& \frac18 (u^t)^2f_0 Mr_0^3 \left[ \partial_t^2\partial_r - 2f_0\partial_t \partial_r^2+f_0^2\partial_r^3 +\frac6{r_0}\partial_t^2 - 2(9-13M/r_0)\partial_t\partial_r + 12\frac{f_0}{r_0}(1-M/r_0)\partial_r^2
\right.\nonumber\\
 &&\phantom{xxxxx} \left.-\frac6{r_0^2}(5-4M/r_0)\partial_t + \frac2{r_0^2}(17-32M/r_0+8M^2/r_0^2)\partial_r 
+\frac{16}{r_0^3}(1-M^2/r_0^2)\right] \sum_m \,\textrm{Re}(\Psi_{\ell m})\,  {}_2\!Y_{\ell m}.	
\eea\label{eq:ar-il}\esube

The mode-sum renormalization of the self-acceleration is given by 
\be
a^{\rm ren\, r} 
 = \lim_{\ell_{\rm max}\rightarrow\infty}\sum_{\ell=0}^{\ell_{\rm max}} (a^{{\rm ret}\, r}[\ell] -a^{{\rm s}\, r}[\ell]), 
\ee
with $a^{{\rm s}\, r}[\ell]$ the singular part of $a^{{\rm ret}\,r}[\ell]$.  As described in paper I, one can identify $a^{\rm s\alpha} [\ell]$ with the 
leading and subleading terms in $L$ in the large-$L$ expansion of $a^{{\rm ret}\alpha}$.   
This expansion has the (direction-dependent) form 
\beq
a^{\rm ret \,r}[\ell] = A^{\pm}\left(\ell+\frac{1}{2}\right) + B + \sum_{k=1}^{\infty}\frac{\tilde E_{2k}}{L^{2k}} 
+ \tilde a_\ell^{\rm ren \, r},
\eeq
where $\tilde a_\ell$ falls off at large $\ell$ faster than any power of $\ell$, and 
the superscript $\pm$ refers to the limit $r\rightarrow r_0^\pm$.  
Because we find that the singular field can be identified with its leading and subleading 
terms, the remaining part of the power-series expansion in $L$ must sum to zero, and 
we reorder the higher-order terms in the power series, replacing $L^k$ by a sequence of polynomials $P_{2k}(\ell)$ that individually sum to zero and whose leading term is $L^k$ 
(see Eq.~(\ref{eq:p2k}), below).  
We find the singular field numerically by matching a power-series of this form, truncated at a finite 
value $k_{\rm max}$ of $k$, to the sequence $a^{\rm ret}[\ell]$, computed using Eqs.~(\ref{eq:ar-l}) and (\ref{eq:ar-il}).
Since the $\ell$-mode expansion of the singular field agrees with the large-$\ell$ expansion of the retarded field, one can extract the regularization parameters by the above method. Though we find the singular field $\psi_0^{\rm s}$ to order O($\epsilon^{-1}$)in the previous paper, one can see that it takes a heavy amount of analytic work to calculate each of the $a^{{\rm s}\,r}_i[\ell]$ which requires a careful analysis of more than 200 terms$!$ Hence, we employ this numerical matching method to extract the renormalization coefficients
in the radiation gauge.

A striking feature of the numerically determined singular field is that the first three 
terms in the $L$ expansion of the conservative part $f^r$ of the self-force coincide 
with the expression for $a^{\rm s \, r}$ obtained from the power-series expansion of 
$\langle \nabla^r\rho^{-1} \rangle$. That is, the average over $\Phi$ of $\partial^r\frac1\rho$ 
is given by 
\be
\langle \partial^r\rho^{-1}\rangle = \sum_\ell [ A^{\pm}\, L\, + B ] + O(\rho),
\ee
with $A^\pm$ and $ B$ given by \cite{dmw03}
\bsube\bea
A^\pm_{\rm analytic} &=& \mp \frac{(1-3M/r_0)^{1/2}}{r_0^2},\\
B_{\rm analytic} &=& -\frac{\left((1-\frac{3M}{r_0})(1-\frac{2M}{r_0})\right)^{1/2}}{r_0^2}
		     \left(F_{1/2}-\frac12\frac{1-3M/r_0}{1-2M/r_0}\, F_{3/2}\right),
\eea\esube
where
\be
  F_p = \frac2\pi\int_0^{\pi/2}\left(1-\frac M{r_0-2M}\sin^2\Phi\right)^{-p} d\Phi.
\ee

Table III shows the agreement between the regularization coefficients that we get in a radiation gauge and the ones in the Lorenz gauge.

\begin{center}
\begin{table}
  \begin{tabular}{|c|c|c|c|c|c|}
    \hline
                 \multicolumn{1}{|p{1.0cm}|}
                 {\centering $r_0/M$}
                & \multicolumn{1}{|p{3.5cm}|}
                 {\centering $A^+$ }  
                & \multicolumn{1}{|p{3.5cm}|}
                 {\centering B} 
                 & \multicolumn{1}{|p{3.0cm}|}
                 {\centering red$\left|\frac{A^+}{A^+_{\rm analytic}}\right|-1$}
                 & \multicolumn{1}{|p{3.0cm}|}
                  {\centering $\left|\frac{B}{B_{\rm analytic}}\right|-1$} \\

    \hline
       6	&	-$1.964185503296099\times 10^{\text{-2}}$	&	-$9.719920769918032\times 10^{\text{-3}}$	&	$6.8 \times 10^{-14}$	&	$-5.0\times 10^{-11}$	\\
       7	&	-$1.542712134731597\times 10^{\text{-2}}$	&	-$7.595781032643107\times 10^{\text{-3}}$	&	$3.7 \times 10^{-14}$	&	$-3.0 \times 10^{-11}$	\\
       8	&	-$1.235264711003273\times 10^{\text{-2}}$	&	-$6.072295959309139\times 10^{\text{-3}}$	&	$ < 
10^{-16}$	&	$-5.7 \times 10^{-12}$	\\
       9	&	-$1.008020470281125\times 10^{\text{-2}}$	&	-$4.954081856693618\times 10^{\text{-3}}$	&	$1.8 \times 10^{-14}$	&	$-3.3 \times 10^{-12}$	\\
       10	&	-$8.366600265340854\times 10^{\text{-3}}$	&	-$4.113353788131433\times 10^{\text{-3}}$	&	$1.2 \times 10^{-14}$	&	$-7.3 \times 10^{-12}$	\\
       11	&	-$7.047957565474786\times 10^{\text{-3}}$	&	-$3.467126055149815\times 10^{\text{-3}}$	&	$8.9 \times 10^{-15}$	&	$-3.7 \times 10^{-12}$	\\
       12	&	-$6.014065304058753\times 10^{\text{-3}}$	&	-$2.960554843842139\times 10^{\text{-3}}$	&	$2.5 \times 10^{-14}$	&	$-1.1 \times 10^{-11}$	\\
       13	&	-$5.189692421934956\times 10^{\text{-3}}$	&	-$2.556541533529994\times 10^{\text{-3}}$	&	$-2.4 \times 10^{-14}$	&	$6.5 \times 10^{-12}$	\\
       14	&	-$4.522475818510165\times 10^{\text{-3}}$	&	-$2.229391187912286\times 10^{\text{-3}}$	&	$8.4 \times 10^{-14}$	&	$-2.9 \times 10^{-11}$	\\
       15	&	-$3.975231959999661\times 10^{\text{-3}}$	&	-$1.960906506358998\times 10^{\text{-3}}$	&	$8.9 \times 10^{-15}$	&	$-3.6 \times 10^{-12}$	\\
       16	&	-$3.521046167445508\times 10^{\text{-3}}$	&	-$1.737934146698723\times 10^{\text{-3}}$	&	$5.8 \times 10^{-14}$	&	$-1.7 \times 10^{-11}$	\\
       17	&	-$3.140087242121197\times 10^{\text{-3}}$	&	-$1.550788700414580\times 10^{\text{-3}}$	&	$-4.6 \times 10^{-15}$	&	$-6.7 \times 10^{-15}$	\\
       18	&	-$2.817502867825028\times 10^{\text{-3}}$	&	-$1.392217662603554\times 10^{\text{-3}}$	&	$3.5 \times 10^{-14}$	&	$-1.1 \times 10^{-11}$	\\
       19	&	-$2.542002591363557\times 10^{\text{-3}}$	&	-$1.256707170227638\times 10^{\text{-3}}$	&	$-3.1 \times 10^{-15}$	&	$9.4 \times 10^{-13}$	\\
       20	&	-$2.304886114323209\times 10^{\text{-3}}$	&	-$1.140007036978271\times 10^{\text{-3}}$	&	$-5.7 \times 10^{-15}$	&	$1.8 \times 10^{-12}$	\\
       25	&	-$1.500933043143495\times 10^{\text{-3}}$	&	-$7.437542878990537\times 10^{\text{-4}}$	&	$-1.6 \times 10^{-15}$	&	$-3.0 \times 10^{-14}$	\\
       30	&	-$1.054092553389464\times 10^{\text{-3}}$	&	-$5.230319186714355\times 10^{\text{-4}}$	&	$4.2 \times 10^{-15}$	&	$-1.1 \times 10^{-12}$	\\
       35	&	-$7.805574591571754\times 10^{\text{-4}}$	&	-$3.876932093890911\times 10^{\text{-4}}$	&	$2.6 \times 10^{-14}$	&	$-6.5\times 10^{-12}$	\\
       40	&	-$6.011057519272318\times 10^{\text{-4}}$	&	-$2.987922074634742\times 10^{\text{-4}}$	&	$3.8 \times 10^{-15}$	&	$-8.5 \times 10^{-13}$	\\
       45	&	-$4.770823620144716\times 10^{\text{-4}}$	&	-$2.372891353438443\times 10^{\text{-4}}$	&	$9.3 \times 10^{-15}$	&	$-2.5 \times 10^{-12}$	\\
       50	&	-$3.878143885933070\times 10^{\text{-4}}$	&	-$1.929854912525034\times 10^{\text{-4}}$	&	$1.8 \times 10^{-15}$	&	$-2.3 \times 10^{-13}$	\\
       55	&	-$3.214363205318354\times 10^{\text{-4}}$	&	-$1.600201924066196\times 10^{\text{-4}}$	&	$6.9 \times 10^{-15}$	&	$-1.7 \times 10^{-12}$	\\
       60	&	-$2.707442873558057\times 10^{\text{-4}}$	&	-$1.348310298624353\times 10^{\text{-4}}$	&	$4.2 \times 10^{-15}$	&	$-9.6 \times 10^{-13}$	\\
       65	&	-$2.311598761671936\times 10^{\text{-4}}$	&	-$1.151520162987173\times 10^{\text{-4}}$	&	$2.0 \times 10^{-16}$	&	$-8.0 \times 10^{-14}$	\\
       70	&	-$1.996605675599292\times 10^{\text{-4}}$	&	-$9.948607781191319\times 10^{\text{-5}}$	&	$-2.0 \times 10^{-16}$	&	$-5.0\times 10^{-14}$	\\
       75	&	-$1.741859372645818\times 10^{\text{-4}}$	&	-$8.681200066822863\times 10^{\text{-5}}$	&	$1.3 \times 10^{-15}$	&	$-3.4 \times 10^{-13}$	\\
       80	&	-$1.532923192995987\times 10^{\text{-4}}$	&	-$7.641384666121239\times 10^{\text{-5}}$	&	$2.7 \times 10^{-15}$	&	$-5.6 \times 10^{-13}$	\\
       85	&	-$1.359438646187131\times 10^{\text{-4}}$	&	-$6.777766234583098\times 10^{\text{-5}}$	&	$-3.0 \times 10^{-16}$	&	$2.9 \times 10^{-13}$	\\

    \hline
  \end{tabular}
  \caption{The first column shows the radius of the orbiting particle in terms of background Schwarzschild coordinate $r$. The second and the third columns show the leading and the sub-leading regularization parameters that we get by a numerical matching. The fourth and the fifth columns show the fractional difference between the numerical and the analytic values.}
  \end{table}
\end{center}

The renormalized self-force is given by $f^r = {\frak m}a^r$, with
\bea
a^{\rm ren \,r} &=& \sum_{\ell=0}^\infty \left(a^{\textrm{ret}\,r}[\ell]- A^\pm L-B \right).
\eea
We can add parts of the perturbation corresponding to a change in mass, and angular momentum in any 
convenient gauge.  They, together with an even-parity $l=1$ gauge transformation outside $r_0$ that 
accounts for a change in the center of mass, have been computed in a Lorenz gauge by Poisson and Detweiler, 
and we use their results.  Note that the singular part of the self-force is given by the coefficients 
in the large-$L$ expansion of the metric and is therefore independent of the choice of gauge for 
$\ell=0$ and $\ell=1$.   Our results for the self-force, expressed as the self-acceleration 
$a_r = f_r/{\frak m}$, are tabulated in Table ~\ref{table3}.   

\begin{center}
\begin{table}
  \begin{tabular}{|c|c|c|c|c|c|}
    \hline
                 \multicolumn{1}{|p{1.0cm}|}
                 {\centering $r_0/M$}
                & \multicolumn{1}{|p{2.2cm}|}
                 {\centering $F^r/\mu^2$} \\

    \hline
       6    &   0.031741815	\\
       7    &   0.024314669	\\
       8    &   0.019541501	\\
       9    &   0.016137918	\\
       10   &   0.013580536	\\
       11   &   0.011595880	\\
       12   &   0.010019806	\\
       13   &   0.0087455255	\\
       14   &   0.0076999148	\\
       15   &   0.0068310918	\\
       16   &   0.0061012423	\\
       18   &   0.0049526422	\\
       20   &   0.0040997900	\\
       25   &   0.0027292140	\\
       30   &   0.0019459393 	\\
       35   &   0.0014569286	\\
       40   &   0.0011313990	\\
       45   &   0.00090387369	\\
       50   &   0.00073864055	\\
       60   &   0.00051979901	\\
       70   &   0.00038553381	\\
       80   &   0.00029728330	\\
       90   &   0.00023619526	\\

    \hline
  \end{tabular}
  \caption{The radial self-force for a particle in circular orbit around a Schwarzschild black hole. The first column lists the Schwarzschild radial coordinate of the orbit; the second gives the radial component of the 
self-force per unit mass square.}
\label{table3} \end{table}
\end{center}
\newpage

\section{Gauge-invariant quantities}
\label{sec4}

The perturbed metric of a particle in circular orbit in a Schwarzschild spacetime is helically 
symmetric: The helical Killing vector $k^\alpha = t^\alpha + \Omega \phi^\alpha$ of the 
background spacetime Lie derives the particle's trajectory and is a Killing vector of the linearized metric
$h^{\rm ret}_{\alpha\beta}$ in a Lorenz, Regge-Wheeler, or radiation gauge.      
As Detweiler noted \cite{detweiler05,detweiler08}, the quantity $H := h_{\alpha\beta}u^\alpha u^\beta/2$ is invariant under gauge transformations generated by a gauge vector $\xi^\alpha$ that shares this symmetry, 
\be
     \Lie_{\bf k}\xi^\alpha = 0.	
\label{eq:liekxi}\ee
A quick covariant derivation follows from the background geodesic equation in the form 
$u^\beta\nabla_\beta u^\alpha = 0$ and the relation $u^\alpha = u^t k^\alpha$, where 
$u^t$ is the scalar $u^\alpha\nabla_\alpha t$.  
The change in $H$ under a gauge transformation is given by 
$\delta_\xi H = u^\alpha u^\beta\nabla_\alpha\xi_\beta $.  Then
\be 
   u^\alpha u^\beta \nabla_\alpha\xi_\beta  = - u^t u^\beta \xi_\alpha \nabla_\beta k^\alpha
		= - u^\beta\xi_\alpha \nabla_\beta u^\alpha= 0,  
\ee
where Eq.~(\ref{eq:liekxi}) is used in the first equality and the geodesic equation, 
together with $k^\alpha \nabla_\alpha (u^t) = 0$, is used to obtain the second.

In this section we describe the computation of the renormalized invariant $H$ and, in 
effect, compare its values at different orbital radii to the values computed in 
two different gauges by Barack and Sago and by Detweiler. As we show in Appendix A of I, 
the  transformation from a Lorenz gauge to the partial radiation gauge we use is generated by a 
helically symmetric gauge vector; the value of $H^{\rm ren}$ in our radiation gauge 
must therefore coincide up to numerical error with its value for a Lorenz gauge. 
The $\ell=0$ part of the gauge transformation from a Lorenz gauge to the Regge-Wheeler gauge, however,
is {\rm not} helically symmetric, and we must take account of the gauge change in $H$ to compare 
our value to the Regge-Wheeler value. 

As we discuss below, instead of $H$ itself, Sago et al. tabulate a related gauge-invariant 
quantity that they term $\Delta U$, which is given in terms of $H$ and the background geometry. 
To facilitate the comparison to their work, we use $\Delta U$ as the quantity to tabulate.  

To compute $H^{\rm ren}$ at the position of the particle, 
we use the harmonic decomposition of the metric to write  
\be
	H^{\rm ret} = \sum_{\ell m} H_{\ell m} Y_{\ell m}(\pi/2,0). 
\ee
We then match the sequence of values, 
\be
  H^{\rm ret}_\ell = \sum_m H_{\ell m} Y_{\ell m}(\pi/2,0), 
\ee
to a power series in $\ell$ of the form 
\beq
E_0 + \frac{E_2}{(\ell-1/2)(\ell+3/2)} + \frac{E_4}{(\ell-3/2)(\ell-1/2)(\ell+3/2)(\ell+5/2)} + \cdots .
\eeq
Because the series is obtained from $H^{\rm ret}$, the renormalization coefficients $E_k$ are again 
invariant under helically symmetric gauge transformations, and we compare in Table \ref{tableiii}  the leading term $E_0$ to an analytic form derived by Detweiler, 
\beq
E_{0\, \textrm{analytic}} 
   = \sqrt{\frac{r-3M}{r^2(r-2M)}}{\,}_2F_1\left(\frac{1}{2},\frac{1}{2},1,\frac{M}{r-2M}\right).
\eeq
 When $H^{\textrm{ret}}$ is decomposed as a mode sum, the information of the singular part is stored in higher $\ell$s. We obtain the leading term by matching $H^{\textrm{ret}}_\ell$ to a power series in $\ell$ from $\ell \simeq 15$ to 85.  The method of matching is explained in Sec.~\ref{sec5} below.

Table~\ref{tableiii} shows the fractional error with which the leading coefficient $E_0$ differs from its analytic form for a set of different radii.

\begin{center}
\begin{table}[h]
  \begin{tabular}{|c|c|c|c|c|c|}
    \hline
                 \multicolumn{1}{|p{1.0cm}|}
                 {\centering $r_0/M$}
                & \multicolumn{1}{|p{2.1cm}|}
                 {\centering $|\Delta E_0/E_0|$ }  \\

    \hline
       6	&	$3.8\times10^{-14}$	\\
       7	&	$1.6\times10^{-14}$	\\
       8	&	$9.0\times10^{-16}$	\\
       9	&	$9.0\times10^{-15}$	\\
       10	&	$1.1\times10^{-15}$	\\
       11	&	$5.6\times10^{-15}$	\\
       12	&	$6.0\times10^{-16}$	\\
       13	&	$2.4\times10^{-15}$	\\
       14	&	$1.1\times10^{-15}$	\\
       15	&	$3.4\times10^{-15}$	\\
       16	&	$2.7\times10^{-15}$	\\
       18	&	$4.0\times10^{-16}$	\\
       20	&	$2.0\times10^{-15}$	\\
       25	&	$2.3\times10^{-15}$	\\
       30	&	$1.4\times10^{-15}$	\\
       40	&	$7.0\times10^{-16}$	\\
       50	&	$1.2\times10^{-15}$	\\
       60	&	$1.7\times10^{-15}$	\\
       70	&	$< 10^{-16}$	\\
       80	&	$5.0\times10^{-16}$	\\
    \hline
  \end{tabular}
  \caption{The first column shows the radius of the orbiting particle in terms of the background Schwarzschild coordinate. The second column shows the fractional error in $E_0$, $\Delta E_0/E_0 = | E_{0\,\textrm{numerical}} - E_{0\,\textrm{analytic}}|/E_{0\,\textrm{analytic}}$.}
 \label{tableiii} \end{table}
\end{center}

In comparing quantities that we compute in a radiation gauge to those computed by Barack and Sago in a Lorenz gauge and by Detweiler in a Regge-Wheeler gauge, we follow the terminology in Sago et al. \cite{sbd08}, using the abbreviations {\em BS} and {\em SD} to refer to quantities computed in the two different gauges.  We have already mentioned that the comparison requires a correction arising from the lack of helical symmetry in the $l=0$ gauge vector, but there is an additional 
difference between the BS and SD computations: 
BS parametrize the perturbed trajectory by proper time $\tau$ with respect to the background metric, 
while SD uses proper time $\widehat\tau$ with respect to the renormalized perturbed metric.
\footnote{To maintain a notation consistent with the EMRI literature and with our previous 
papers, we denote by $u^\alpha$ the four-velocity normalized with respect to the background metric, with 
$\tau$ the corresponding background proper time.  Sago et al. use $\tilde u^\alpha$ and $\tilde \tau$ 
for these quantities and their quantities with no tilde correspond to proper time with respect to 
the perturbed metric.} 
From the relations  
\be
  \widehat u^\alpha = u^\alpha \frac{d\tau}{d\widehat \tau}, \qquad 
(g_{\alpha\beta}+h^{\rm ren}_{\alpha\beta})\widehat u^\alpha \widehat u^\beta = 1 
		= g_{\alpha\beta} u^\alpha u^\beta,
\ee 
we have, to linear order in the perturbation, 
\beq
\frac{d\tau}{d\widehat\tau} = 1 - \frac{1}{2}h^{\rm ren}_{\alpha\beta} u^\alpha u^\beta.
\eeq

Because we, like BS, use proper time with respect to the background metric, we have chosen a Lorenz gauge 
for the $l=0$ and $l=1$ parts of the perturbation so that our $u^t$ will coincide with that of BS.
In comparing to SD, we then need both corrections -- from the $l=0$ gauge transformation and from
the reparametrization of the trajectory, as given in Sago et al. \cite{sbd08}. 
The $\ell=0$ gauge transformation has gauge vector $\xi^\alpha = \alpha t\, t^\alpha$, 
with 
\be
	\alpha = \mathfrak{m}/\sqrt{r_0(r_0-3M)}.
\ee
The relation between $u^t$ in our 
modified radiation gauge and in a Lorenz gauge (normalized by proper time with respect 
to the background metric) and $\widehat u^t$ in a Regge-Wheeler 
gauge (normalized by proper time with respect to the perturbed metric) is then 
\beq
\widehat u^t = u^t(1 + \alpha - H^{\rm ren}) + O(\mathfrak{m}^2),
\eeq
where $H^{\rm ren}$, given by
\bea
  H^{\rm ren} &=& \frac{1}{2}h^{\rm ren}_{\alpha\beta} u^\alpha u^\beta,
\eea
is computed in our modified radiation gauge. Because $H^{\rm ret}$ is invariant 
under helically symmetric gauge transformations, $H^{\rm ren}$ is similarly 
invariant.

Following Sago et al., we write $U=\widehat u^t$ and construct from it another  
quantity that is invariant under helically symmetric gauge transformations 
by expressing the unperturbed $U$ as a function 
of $\Omega$: With  
\be
   {\cal U}_0(\Omega) := (1-3M/R)^{-1/2}, \quad R := \left(\frac{M}{\Omega^2}\right)^{1/3}, 
\label{eq:deltaOmega}\ee
the unperturbed values $U_0$ and $\Omega_0$ satisfy 
\be
        U_0 = {\cal U}_0(\Omega_0).
\label{eq:U0}\ee
Then $\Delta U$ is defined as the difference  
\bea
\Delta U := U - {\cal U}_0(\Omega)= U- (1-3M/R)^{-1/2}, 
\label{eq:DeltaU}\eea
where the angular velocity that appears in ${\cal U}_0(\Omega)$ is the angular velocity of 
the perturbed trajectory.    
One can express $\Delta U$ in terms of $H$ using the changes $\delta\Omega$ and $\delta U$ 
in $\Omega$ and $U$ due to a self-force whose only component is radial, $f_r={\frak m} a_r$, 
namely      
\bea
\delta\Omega &=& \Omega -\Omega_0 = -\Omega_0  \frac{r_0^2(1-3M/r_0)}{2M}a_r + O(\mathfrak{m}^2), \\
\delta U &=& U-U_0 =  U_0 (-H^{\rm ren}-\frac{r_0}2 a_r) +  O(\mathfrak{m}^2), 
\eea
to obtain
\be
\Delta U = -(1-3M/r_0)^{-1/2}H^{\rm ren}.
\ee

Note that $\Delta U$ is gauge-invariant in the standard sense that it has the same value when 
$h_{\alpha\beta}$ is replaced by $h_{\alpha\beta} + \Lie_\xi g_{\alpha\beta}$ (in this case, 
when $\xi^\alpha$ is helically symmetric).  
Detweiler and Sago et al. \cite{detweiler05,detweiler08,sbd08} also use the term ``gauge-invariant'' 
to refer to the finite quantities $U$ and $\Omega$.  Their terminology is motivated by the fact 
that $\Omega$ and $U$ are physically meaningful: In particular, $\Omega$ 
can be measured by an observer at infinity from the periodicity of received signals sent at 
equally spaced proper times from the orbiting particle.  

One must be careful, however, because $\delta \Omega$ and $\delta U$ are not gauge invariant.  
With the standard definition of the change of a scalar on spacetime under a gauge transformation 
generated by a gauge vector $\xi^\alpha$, they change by 
\be
\delta \Omega \rightarrow \delta \Omega + \Lie_\xi\Omega, \qquad \delta U\rightarrow \delta U+\Lie_\xi U.  
\ee 
Note, in particular, that Sago et al. use symbols $\delta_\xi \Omega$ and $\delta_\xi U$ 
whose meaning is different from the change in $\Omega$ and $U$ under 
a gauge transformation, while their symbol $\delta_\xi h_{\alpha\beta}$ {\em does}
have the meaning $\Lie_\xi g_{\alpha\beta}$.      
We discuss the terminology and the reason $\Delta U$ is invariant under helically symmetric gauge 
transformations in the Appendix.

Because we have used the Lorenz gauge for the $\ell=0$ and $\ell=1$ parts of the metric 
perturbation (derived by Detweiler and Poisson), 
and the gauge transformation from a Lorenz gauge to the Regge-Wheeler gauge of Detweiler is not helically symmetric, 
we need to make the same gauge adjustment made by BS to our $\Delta U$ to compare its the value 
to that of Detweiler, namely
\be
\Delta U^{\rm D} = (1-3M/r_0)^{-1/2}\left( \frac{r_0-2M}{r_0-3M}\alpha - H^{\rm ren,RG}\right).
\ee

In the comparison table, Table~\ref{table5} below, $\Delta U$ is given in the Regge-Wheeler gauge used by 
Detweiler. The table reports results of the computation performed in the three gauges, 
radiation, Regge-Wheeler, and Lorenz.  
\begin{center}
\begin{table}[h]
  \begin{tabular}{|c|c|c|c|}
    \hline
                 \multicolumn{1}{|p{1.0cm}|}
                 {\centering $r_0/M$}
                & \multicolumn{1}{|p{2.2cm}|}
                 {\centering $\Delta U$} 
                & \multicolumn{1}{|p{2.1cm}|}
                 {\centering $\Delta U(\textrm{from SD})$}
                & \multicolumn{1}{|p{2.2cm}|}
                 {\centering $\Delta U(\textrm{from BS})$}  \\

    \hline
       6    &   -0.29602751   &   -0.2960275	&	-0.296040244	\\
       7    &   -0.22084753   &   -0.2208475	&	-0.220852781	\\
       8    &   -0.17771974   &   -0.1777197	&	-0.177722443	\\
       9    &   -0.14936061   &   -0.1493606	&	-0.149362192	\\
       10   &   -0.12912227   &   -0.1291222	&	-0.129123253	\\
       11   &   -0.11387465   &   -0.1138747	&	-0.113875315	\\
       12   &   -0.10193557   &   -0.1019355	&	-0.101936046	\\
       13   &   -0.092313311  &   -0.09231331	&	-0.092313661	\\
       14   &   -0.084381953  &   -0.08438195	&	-0.084382221	\\
       15   &   -0.077725319  &   -0.07772532	&	-0.077725527	\\
       16   &   -0.072055057  &   -0.07205505	&	-0.072055223	\\
       18   &   -0.062901899  &   -0.06290189	&	-0.062902026	\\
       20   &   -0.055827719  &   -0.05582771	&	-0.055827795	\\
       25   &   -0.043599843  &   -0.04359984	&	-0.043599881	\\
       30   &   -0.035778314  &   -0.03577831	&	-0.035778334	\\
       40   &   -0.026339677  &   -0.02633967	&	-0.026339690	\\
       50   &   -0.020844656  &   -0.02084465 	&	-0.020844661 	\\
       60   &   -0.017247593  &   -0.01724759	&	-0.017247596	\\
       70   &   -0.014709646  &   -0.01470964	&	-0.014709648	\\
       80   &   -0.012822961  &   -0.01282296	&	-0.012822962	\\
       90   &   -0.011365316  &   -0.01136531	&	-0.011365317	\\

    \hline
  \end{tabular}
  \caption{In this table we compare our values of $\Delta U$ with BS and SD. The first column shows the radius of the orbiting particle in terms of background Schwarzschild coordinate. The second, third and fourth column shows the values of $\Delta U$ computed in a radiation gauge, Regge-Wheeler gauge (SD) and Lorenz gauge (BS).}
\label{table5}\end{table}
\end{center}

\newpage
\section{Numerical matching}
\label{sec5}
As mentioned in Paper I, we find the singular part of the self-force by matching a power series to its numerically computed large-$L$ behavior.   Explicitly, we match the sequence of 
values $a^{\rm ret}[\ell]$ to successive terms in a series of the form 
\beq
A L + B + \sum_{k=1}^{k_{\rm max}}\frac{\tilde E_{2k}}{P_{2k}(\ell)},
\label{eq:e2k}\eeq
where the polynomial, 
\be
P_{2k}(\ell) = \prod_{i=1}^k(\ell-k-\frac{1}{2}+i)\prod_{j=1}^k(\ell+k+\frac{3}{2}-j),
\label{eq:p2k}\ee
satisfies 
\be \label{sumtozero}
\sum_{\ell=0}^\infty \frac{1}{P_{2k}(\ell)} = 0.
\ee

When one matches the above series to the numerical series  $a^{\rm ret}[\ell]$ over a range $\ell_{\textrm{min}} \leq\ell\leq \ell_{\textrm{max}}$, the accuracy with which $A$ and $B$ are are obtained depends on the value of 
$k_{\rm max}$ chosen -- on the number of parameters $E_{2k}$ used in the matching.  We choose $k_{\rm max}$ to 
minimize the difference $(A_{(k_{\rm max}+1)}-A_{k_{\rm max}})/A_{k_{\rm max}}$ as a function of $k_{\rm max}$.
For $\ell_{\rm max}$ infinite and no numerical error,  $A_{(k_{\rm max}+1)}-A_{k_{\rm max}}$ would converge 
to zero as $k_{\rm max}\rightarrow \infty$.  For finite $\ell_{\textrm{max}}$ only a finite number of 
parameters can be extracted, and we approximate the value of $k_{\textrm{max}}$ for which $A$ and $B$ are most 
accurately determined by the value $\tilde{k}$ of $k_{\textrm{max}}$ that minimizes 
$A_{(k_{\rm max}+1)}-A_{k_{\rm max}}$.  

\begin{figure}[h]
\includegraphics[width = 6in]{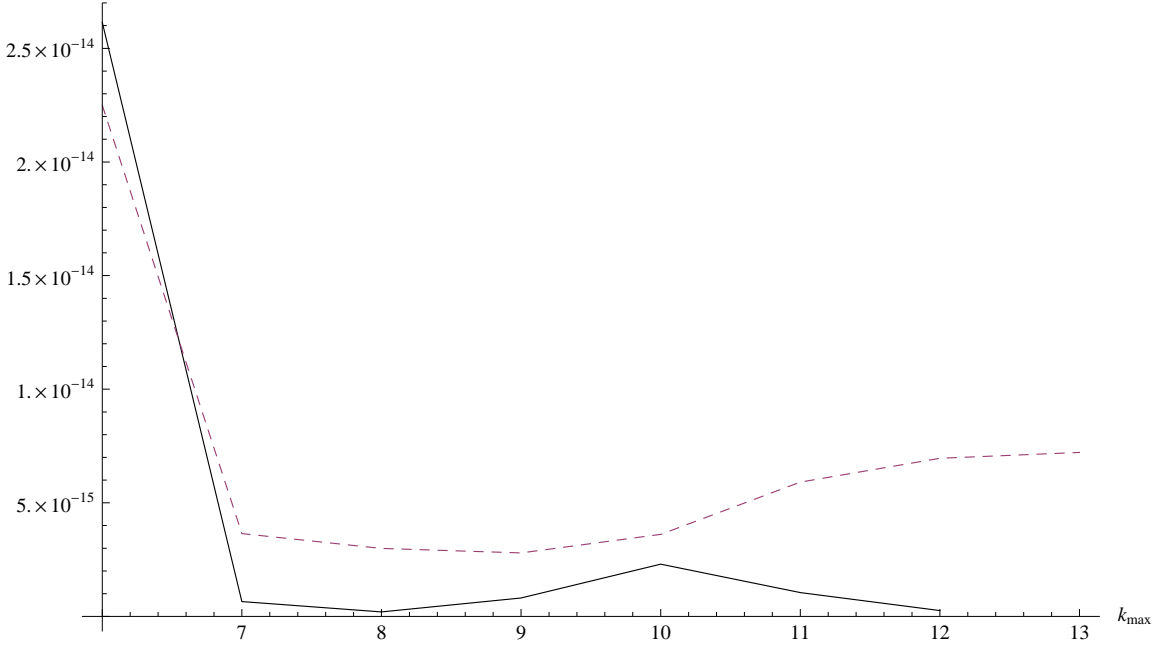}
\caption{The solid curve is a plot of $|E_{0(k_{\rm max}+1)}-E_{0 k_{\rm max}}|/E_{0 k_{\rm max}}$ vs $k_{\rm max}$. The dashed curve is a plot of $|E_{0 analytic}-E_{0 \tilde{k}}|/(E_{0 analytic})$ as a function of $k_{\rm max}$. Both are for orbital radius $r_0=10M$}
\label{fig1}
\end{figure}

We check this approximation by using our knowledge of the analytic form of the leading renormalization 
parameter $E_0$ in the quantity $h_{\alpha\beta}u^\alpha u^\beta = 2H$:
\beq
2H^\textrm{sing} = E_0 + \sum_{k=1}^{k_{\rm max}} \frac{E_{2k}}{P_{2k}(\ell)}.
\eeq
In Fig. \ref{fig1}, we compare $\tilde k$ to the value of $k_{\rm max}$ that minimizes the error in $E_0$.
The graph shows for a particular orbital radius that the error in $E_0$ is a minimum at a value of 
$k_{\rm max}$ in the interval $\tilde{k}\pm 1$. 

Table $\ref{tablevii}$ shows for a set of orbital radii the minimum value of $|E_{0 (k_{\rm max}+1)}-E_{0 k_{\rm max}}|/E_{0 k_{\rm max}}$, its value at $k_{\rm max}=\tilde k$; that is compared to the value of 
$|E_{0 analytic}-E_{0 k_{\rm max}}|/E_{0 analytic}$, again at $k_{\rm max}=\tilde k$.   One can infer from the table that, by minimizing $A_{(k_{\rm max}+1)}-A_{k_{\rm max}}$, one obtains values of the renormalization parameters to a 
fractional precision of $10^{-13}$ or better.  Similar accuracy was reported in Paper I for the leading renormalization parameter in the axisymmetric part of $\psi_0$. 
\newpage

\begin{center}
\begin{table} [h]
  \begin{tabular}{|c|c|c|c|c|c|}
    \hline
                 \multicolumn{1}{|p{1.0cm}|}
                 {\centering $r_0/M$}
                & \multicolumn{1}{|p{5.0cm}|}
                 {\centering minimum of $\frac{ E_{0 (k_{\rm max}+1)} - E_{0 k_{\rm max}} }{ E_{0 k_{\rm max}} }$ }  
                & \multicolumn{1}{|p{5.0cm}|}
                 {\centering $\frac{E_{0 analytic}-E_{0 \tilde{k}}}{E_{0 analytic}}$}  \\

    \hline
       6	&	$1.4\times10^{-14}$	&	$1.4\times10^{-14}$	\\
       7	&	$1.5\times10^{-15}$	&	$1.1\times10^{-14}$	\\
       8	&	$1.7\times10^{-15}$	&	$1.3\times10^{-15}$	\\
       9	&	$3.0\times10^{-16}$	&	$8.4\times10^{-15}$	\\
       10	&	$2.0\times10^{-16}$	&	$3.0\times10^{-15}$	\\
       11	&	$8.0\times10^{-16}$	&	$3.8\times10^{-15}$	\\
       12	&	$< 10^{-16}$	&	$2.7\times10^{-15}$	\\
       13	&	$3.0\times10^{-16}$	&	$5.0\times10^{-16}$	\\
       14	&	$2.0\times10^{-16}$	&	$5.1\times10^{-15}$	\\
       15	&	$2.0\times10^{-16}$	&	$9.0\times10^{-16}$	\\
       16	&	$< 10^{-16}$	&	$1.7\times10^{-15}$	\\
       18	&	$2.0\times10^{-16}$	&	$2.0\times10^{-16}$	\\
       20	&	$< 10^{-16}$	&	$2.0\times10^{-15}$	\\
       25	&	$< 10^{-16}$	&	$2.5\times10^{-15}$	\\
       30	&	$4.0\times10^{-16}$	&	$6.0\times10^{-16}$	\\
       40	&	$3.0\times10^{-16}$	&	$5.0\times10^{-16}$	\\
       50	&	$2.0\times10^{-16}$	&	$1.1\times10^{-15}$	\\
       60	&	$6.0\times10^{-16}$	&	$1.4\times10^{-15}$	\\
       70	&	$< 10^{-16}$	&	$3.0\times10^{-16}$	\\
       80	&	$2.0\times10^{-16}$	&	$< 10^{-16}$	\\
    \hline
  \end{tabular}
  \caption{The first column shows the radius of the orbiting particle in terms of background Schwarzschild coordinate $r$. The second column shows the minimum of $|E_{0(k_{\rm max}+1)}-E_{0 k_{\rm max}}|/E_{0 k_{\rm max}}$ as a function of $k_{\rm max}$. Let that $k_{\rm max}$ be denoted as $\tilde{k}$. The third column shows the quantity $|E_{0 analytic}-E_{0 \tilde{k}}|/(E_{0 analytic})$. $\ell_{\textrm{min}}$ = 15. }
 \label{tablevii} \end{table}
\end{center}

The accuracy with which the values of $E_k$ are obtained also depends on the values of $\ell_{\textrm{min}}$ 
and $\ell_{\textrm{max}}$. For fixed $\ell_\textrm{max}$, there is an optimal value of 
$\ell_{\textrm{min}}$ that minimizes the error in the regularization parameters.  
For higher $\ell_{\textrm{min}}$, contributions of the higher-order $E_{2k}$ are too small to 
extract; for small $\ell_{\textrm{min}}$ the $H^{\rm ret}[\ell]$ (or $a^{{\rm ret}\, r}[\ell]$) 
will depart from its large-$L$ behavior.  Tables \ref{awesometable1} and \ref{awesometable2} show the behavior of the regularization parameters for different choice of $\ell_{min}$ (for a fixed $\ell_{\rm max} = 84$) at two different radii.  Increasing $\ell_{\textrm{max}}$, of course, increases the 
accuracy of the computation.   
\\

\begin{center}
\begin{table} [h]
  \begin{tabular}{|c|c|c|c|c|c|c|c|}
    \hline
                 \multicolumn{1}{|p{1.0cm}|}
                 {\centering $\ell_{\textrm{min}}$}
                & \multicolumn{1}{|p{1.0cm}|}
                 {\centering $0$ }  
                & \multicolumn{1}{|p{1.0cm}|}
                 {\centering $-2$ }
                & \multicolumn{1}{|p{1.0cm}|}
                 {\centering $-4$ } 
                & \multicolumn{1}{|p{1.0cm}|}
                 {\centering $-6$ } 
                & \multicolumn{1}{|p{1.0cm}|}
                 {\centering $-8$ } 
                 & \multicolumn{1}{|p{3.0cm}|}
                 {\centering $\frac{E_{0 analytical} - E_{0 \tilde{k}}}{E_{0analytical}}$ }
                 & \multicolumn{1}{|p{1.0cm}|}
                 {\centering $\tilde{k}$ }\\
    \hline
    5	&	$5\times10^{-13}$	&	$1\times10^{-9}$	&	$9\times10^{-7}$	&	$4\times10^{-4}$	&	$1\times10^{-3}$	&	$6\times10^{-13}$	&	10	\\
    10	&	$4\times10^{-17}$	&	$8\times10^{-13}$	&	$2\times10^{-9}$	&	$4\times10^{-6}$	&	$5\times10^{-5}$	&	$7\times10^{-16}$	&	11	\\
    15	&	$2\times10^{-16}$	&	$7\times10^{-12}$	&	$4\times10^{-8}$	&	$1\times10^{-4}$	&	$4\times10^{-3}$	&	$8\times10^{-16}$	&	10	\\
    20	&	$2\times10^{-15}$	&	$9\times10^{-11}$	&	$6\times10^{-7}$	&	$3\times10^{-3}$	&	$1\times10^{-1}$	&	$2\times10^{-15}$	&	7	\\
    25	&	$7\times10^{-16}$	&	$4\times10^{-11}$	&	$3\times10^{-7}$	&	$1\times10^{-3}$	&	$6\times10^{-2}$	&	$2\times10^{-15}$	&	6	\\
    \hline
  \end{tabular}
  \caption{This table shows the first five fractional differences between successive regularization parameters in the singular part of $2H$ for five different values of $\ell_{\textrm{min}}$, with
$k_{\rm max} =\tilde{k}$. The second, third, fourth, fifth and the sixth columns list the fractional differences($|E_{n,\tilde{k}+1} - E_{n,\tilde{k}}|/E_{n,\tilde{k}}$) for n = 0,-2,-4,-6,-8, respectively. The seventh column gives the fractional difference between $E_{0 analytical}$ and $E_{0 \tilde{k}}$. The last column gives $\tilde{k}$. All values are for orbital radius $r_0 =15M$.}
  \label{awesometable1}
  \end{table}
\end{center}

\begin{center}
\begin{table}[h]
  \begin{tabular}{|c|c|c|c|c|c|c|c|}
    \hline
                 \multicolumn{1}{|p{1.0cm}|}
                 {\centering $\ell_{\textrm{min}}$}
                & \multicolumn{1}{|p{1.0cm}|}
                 {\centering $0$ }  
                & \multicolumn{1}{|p{1.0cm}|}
                 {\centering $-2$ }
                & \multicolumn{1}{|p{1.0cm}|}
                 {\centering $-4$ } 
                & \multicolumn{1}{|p{1.0cm}|}
                 {\centering $-6$ } 
                & \multicolumn{1}{|p{1.0cm}|}
                 {\centering $-8$ } 
                 & \multicolumn{1}{|p{3.0cm}|}
                 {\centering $\frac{E_{0 analytical} - E_{0 \tilde{k}}}{E_{0analytical}}$ }
                 & \multicolumn{1}{|p{1.0cm}|}
                 {\centering $\tilde{k}$ }\\
    \hline
    5	&	$2\times10^{-13}$	&	$9\times10^{-9}$	&	$4\times10^{-6}$	&	$7\times10^{-3}$	&	$2\times10^{-3}$	&	$7\times10^{-12}$	&	10	\\
    10	&	$6\times10^{-18}$	&	$9\times10^{-14}$	&	$1\times10^{-10}$	&	$1\times10^{-6}$	&	$2\times10^{-6}$	&	$3\times10^{-15}$	&	11	\\
    15	&	$2\times10^{-16}$	&	$6\times10^{-12}$	&	$2\times10^{-8}$	&	$4\times10^{-4}$	&	$2\times10^{-3}$	&	$3\times10^{-15}$	&	8	\\
    20	&	$2\times10^{-16}$	&	$7\times10^{-12}$	&	$4\times10^{-8}$	&	$2\times10^{-3}$	&	$2\times10^{-2}$	&	$3\times10^{-16}$	&	9	\\
    25	&	$1\times10^{-15}$	&	$5\times10^{-11}$	&	$3\times10^{-7}$	&	$9\times10^{-3}$	&	$7\times10^{-2}$	&	$2\times10^{-15}$	&	7	\\
    \hline
  \end{tabular}
  \caption{Entries are as in Table \ref{awesometable1}, with $r_0$=10M. }
  \label{awesometable2}
  \end{table}
\end{center}

The number of renormalization parameters needed to attain machine accuracy is small, because the 
errors in the mode sums are of order
\bea
\sum_{\ell=85}^{\infty}\frac{1}{P_{10}(\ell)} &\sim & \sum_{\ell=84}^{\infty}\frac{1}{(\ell+1/2)^{10}}\sim 10^{-19},\\ \nonumber
\sum_{\ell=85}^{\infty}\frac{1}{P_{8}(\ell)} &\sim & \sum_{\ell=84}^{\infty}\frac{1}{(\ell+1/2)^{8}}\sim 10^{-15}
\eea
and the $|\textrm{reg parameter}| < 1$.
We work to a fractional error of order  $10^{13}$, terminating the sum over $k$ in Eq.~(\ref{4}) at $k_\textrm{max} = 3$.

\begin{figure}[h]
\includegraphics[width = 6in]{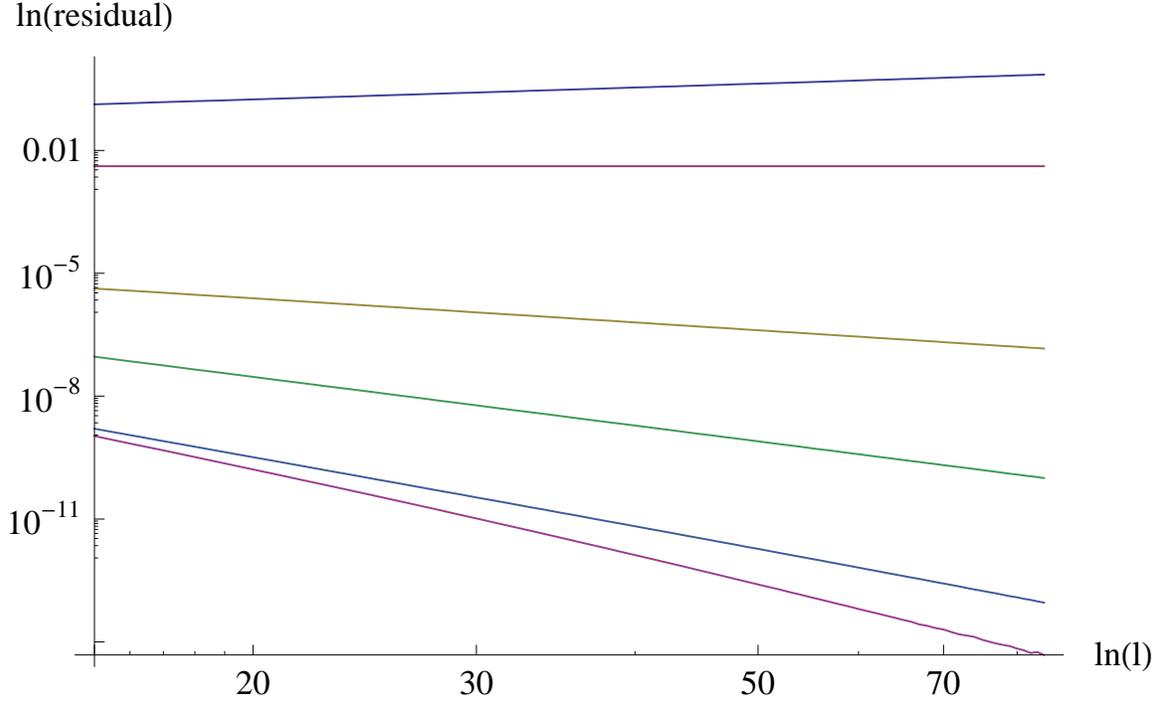}
\caption{This figure shows a plot of $\log a^\textrm{ret r}_{\ell}$ (and subsequent subtractions of the singular terms) vs $\log \ell$.  The topmost curve is $a^\textrm{ret r}_\ell \sim \ell$. The second curve (from the top) is $(a^\textrm{ret r}_\ell - A (\ell+1/2)) \sim \ell^0$. The third curve shows $(a^\textrm{ret r}_\ell - A (\ell+1/2) - B) \sim \ell^{-2}$. The fourth, fifth and sixth curves show the subsequent cumulative subtractions of $E_{2}/P_{2}(\ell)$, $E_{4}/P_{4}(\ell)$ and $E_{6}/P_{6}(\ell)$ from the third. }
\label{fig2}
\end{figure}

Figure~\ref{fig2} shows the result of subtracting successive terms in the numerically-determined 
large-$L$ expansion of the self-force, using $a^{\rm ret\,r}=f^{\rm ret\,r}/\frak{m}$, for 
a particle at $r_0=10M $. 

\newpage

Each successive subtraction should reduce the slope of the 
$\log \ell$ vs $\log f^{\rm ret\,r}[\ell]$ curve by an integer, and the computed graphs 
shows the expected behavior for the first four renormalization coefficients.  

The accuracy with which the higher-order renormalization coefficients can be computed 
can be used to estimate the accuracy with which $H^{\rm ren}$ and $f^{{\rm ren}\,r}$ 
are computed.  In addition, once the $E_{2k}$ have been found, the terms in the 
mode sum can be grouped as a rapidly convergent numerical sum and an analytically 
known sum.  We have 
\bea \label{4}
f^{\textrm{ren},r} &=&  \sum_{\ell=0}^\infty\left[ f^{\textrm{ret}\,r}_\ell - A(\ell+1/2) - B \right]\nonumber \\ 
&=& \left(\sum_{\ell=0}^{\ell_\textrm{max}}+\sum_{\ell_\textrm{max}+1}^{\infty}\right) \left[ f^{\textrm{ret}\,r}_\ell - A (\ell+1/2) - B -\frac{\tilde{E}_2}{Q_2(\ell)} - \cdots - \frac{\tilde{E}_{2k_\textrm{max}}}{Q_{2k_\text{max}}(\ell)} \right] 
 + \sum_{\ell=0}^\infty \left[ \frac{\tilde{E}_2}{Q_2(\ell)} + \cdots + \frac{\tilde{E}_{2k_\textrm{max}}} {Q_{2k_\textrm{max}}(\ell)} \right] 
\nonumber\\ 
&=& \sum_{\ell=0}^{\ell_\textrm{max}} \left[ f^{\textrm{ret}\,r}_\ell - A (\ell+1/2) - B -\frac{\tilde{E}_2}{Q_2(\ell)}  - \cdots - \frac{\tilde{E}_{2k_\textrm{max}}}{Q_{2k_\text{max}}(\ell)}\right] 
+ \sum_{\ell=0}^\infty \left[ \frac{\tilde{E}_2}{Q_2(\ell)} + \cdots + \frac{\tilde{E}_{2k_\textrm{max}}}{Q_{2k_\textrm{max}}(\ell)} \right] + O\left( \frac{1}{\ell_\textrm{max}^{2k_\textrm{max}+1}}\right),\qquad   
\label{eq:modesum}\eea 
where the terms $Q_{2k}(\ell)$ represent any polynomials in $\ell$ whose first term is $\ell^{2k}$.
The polynomials can be chosen to allow the second sum on the right side of Eq.~(\ref{eq:modesum}) 
to be computed analytically.  When the polynomials ${\cal P}_{2k}$ can be used, as in our calculation, the second sum vanishes.  

\section{Discussion}
	The work here gives a first example of the successful use of a (modified) radiation gauge to compute the self-force on a particle in circular orbit. 
In comparing the renormalized perturbed metric component $h_{\alpha\beta}u^\alpha u^\beta$,
we obtain agreement to high numerical accuracy with previous calculations 
in other gauges.  The work shows that the singular field can be identified with 
the singular field in a Lorenz gauge. We verify the coincidence numerically to high precision. It follows analytically from (1) the expression derived in the companion paper \cite{sf2} for the gauge vector $\xi^\alpha$ that relates a Lorenz and radiation gauge 
and (2) the fact that the gauge transformation of the self-force for a particle 
in circular orbit involves no derivatives of the gauge vector.      
 
In extending the method to circular orbits in Kerr and then to more general orbits, additional subtleties arise.  Although spheroidal harmonics decouple in 
the Teukolsky equation for $\psi_0$ and $\psi_4$, the lack of spherical symmetry 
means that angular harmonics with different values of $\ell$ no longer 
decouple in the perturbed metric or in the expression for the self-force.
This significantly changes the way one computes the contribution to the 
perturbed metric associated with the change in the center of mass -- a perturbation
that is purely $\ell=1$ for a Schwarzschild background. A second complication 
for non-circular orbits, arises from the existence of a region between periastron 
and apastron, a region where the time and angular harmonics of 
$\psi_0$ have a nonzero source. Ways to handle each of these complications 
will be discussed in a subsequent paper.

\begin{acknowledgments}
We thank Norichika Sago for corrections and helpful comments and 
Eirini Messaritaki for a number of discussions early in the 
course of the work on this paper and Paper I. 
This work was supported in part by NSF Grants PHY 0503366 and PHY 1001515.  
D.H-K's work was supported by the Alexander
von Humboldt Foundation's Sofja Kovalevskaja Programme funded by the
German Federal Ministry of Education and Research and by WCU (World
Class University) program of NRF/MEST (R32-2009-000-10130-0).
\end{acknowledgments}

\appendix
\section{Gauge invariants}
\label{ap:gaugeinvariants}

As we mentioned in Sec.~\ref{sec4}, the symbols $\delta_\xi \Omega$ and $\delta_\xi U$ used in Refs.~\cite{detweiler05,detweiler08,sbd08} in describing $U$ and $\Omega$ as gauge-invariant are not the standard definitions of the change in a scalar function under a gauge transformation. 
The difference in definitions arises because 
the four-velocity of a single particle does not conform to the framework used to describe 
perturbations of a set of physical fields on spacetime.  
The change in the trajectory of a single particle involves a four-velocity that is 
defined only on a single trajectory, and the trajectory is different for the perturbed 
and unperturbed spacetimes.  

Note that Detweiler and Sago et al. are careful to distinguish a lack of gauge invariance 
of $\delta\Omega:=\Omega -\Omega_0$ and $\delta U:=U-U_0$  from the invariance of $\Omega$ and $U$.  The number $\Omega$ can be measured, for example, by an observer at infinity 
from waves received from the orbiting particle: It is in this sense gauge-invariant. 
However, when $\Omega$ is regarded as a function of trajectories or as a 
function on a Schwarzschild spacetime that assigns to each point of spacetime 
the frequency of a circular orbit through that point, it becomes gauge-dependent.  
In particular, in the active description of a gauge transformation, 
an infinitesimal diffeo changes the metric at a point of spacetime by 
$\Lie_{\bf\xi} g_{\alpha\beta}$, and the frequency of a geodesic through that 
point then changes by $\Lie_\xi\Omega$.
(In the passive description of a gauge transformation, with $\Omega$ regarded 
as a function of the coordinates, an infinitesimal coordinate transformation 
correspondingly changes the functional dependence of $\Omega$ on the coordinates.)            

We give here a brief review of general relativistic perturbation theory and gauge transformations.
We describe linearized perturbation in terms of a family of fields, with 
gauge transformations described in the active sense of diffeos (diffeomorphisms).
We begin with a family of metrics $g_{\alpha\beta}(\lambda)$, defining $\delta g_{\alpha\beta}$  
as the change in the metric to first order in $\lambda$ as the quantity 
\be
  \delta g_{\alpha\beta} = \left.\frac d{d\lambda}g_{\alpha\beta}(\lambda)\right|_{\lambda=0}. 
\ee
When no other fields are present, the metric $g_{\alpha\beta}$ is physically 
equivalent to the diffeomorphically related metric 
$\psi^* g_{\alpha\beta}$, 
where $\psi$ is a diffeo (a diffeomorphism) and $\psi^*$ is the pullback, 
whose components are given by
\be
   \psi^* g_{\mu\nu}(P) =  \frac{\partial\psi^\sigma}{\partial x^\mu}  
			\frac{\partial\psi^\tau}{\partial x^\nu} g_{\sigma\tau}[\psi(P)].
\ee
A gauge transformation is defined by considering a smooth family of diffeos $\psi_\lambda$,
with $\psi_0$ the identity.  We denote by $\xi^\alpha$ the vector field tangent at 
each point $P$ to the orbit $\lambda\mapsto \psi_\lambda(P)$ of $P$.  That is,
to linear order in $\lambda$, a point with coordinates $x^\mu$ is mapped to a point with 
coordinates $\psi^\mu(P)=x^\mu+\lambda\xi^\mu(P)$. 
The family of metrics $g_{\alpha\beta}(\lambda)$ 
is then physically equivalent to the family $\psi^*_\lambda g_{\alpha\beta}(\lambda)$, 
and the metric perturbation $\delta g_{\alpha\beta}$ is physically equivalent to the 
gauge-related metric perturbation 
\be
  \delta g_{\alpha\beta} +\delta_{\bm\xi}g_{\alpha\beta} := \left.\frac d{d\lambda}\psi^*_\lambda g_{\alpha\beta}(\lambda)\right|_{\lambda=0} = \left.\frac d{d\lambda} g_{\alpha\beta}(\lambda)\right|_{\lambda=0}+\left.\frac d{d\lambda}\psi^*_\lambda g_{\alpha\beta}(0)\right|_{\lambda=0}
 = \delta g_{\alpha\beta} +\Lie_{\bm\xi}g_{\alpha\beta}.
\ee  

When other physical fields $[T_1, \ldots, T_N]$ are present,  
the families of fields $[g_{\alpha\beta}(\lambda),T_1(\lambda), \ldots, T_N(\lambda)]$ and 
$[\psi^*_{\lambda} g_{\alpha\beta}, \psi^*_{\lambda} T_1, \ldots, \psi^*_{\lambda} T_N]$ 
describe the same physical system. The corresponding physical 
equivalence of the perturbed fields, linearized about $\lambda=0$, is the equivalence of $\delta T$ 
and $\delta T + \Lie_\xi T$, for each physical field $T$, where    
\be
  \delta T = \left. \frac d{d\lambda} T(\lambda)\right|_{\lambda = 0}, \qquad 
 \delta\,\psi^*_{\lambda} T = \left. \frac d{d\lambda} [\psi^*_{\lambda} T(\lambda)]\right|_{\lambda = 0} 
	= \delta T + \delta_\xi T, \mbox{ and } \delta_{\bm\xi} T = \Lie_{\bm\xi} T.
\label{eq:deltaT}\ee  
In particular, if $T$ is a scalar, $\psi_\lambda^* T = T\circ\psi_\lambda$, implying 
\be
     T(P) = \psi_\lambda^*T [ \psi_\lambda^{-1} (P)].  
\ee

A scalar $f$ constructed locally from a set of tensor fields $T_1, \ldots T_k$, with 
\be
  f = F(T_1, \ldots T_k), \qquad\mbox{with } f(P) = F[T_1(P), \ldots T_k(P)], 
\ee
satisfies 
\be
  \psi f = F(\psi T_1, \ldots \psi T_k), 
\ee
implying, for any vector field $\xi^\alpha$, 
\be
  \Lie_\xi f = \left.\frac d{d\lambda} F(T_1+\lambda\Lie_\xi T_1, \ldots, T_k+\lambda\Lie_\xi T_k)\right|_{\lambda=0}. 
\ee
 
One also uses a set of reference fields -- basis vectors or coordinates, for example --  that 
are independent of $\lambda$ and are not changed by the diffeo that maps the set of physical 
fields to the physically equivalent set.  For example, with $t^\alpha$ the vector $\bm\partial_t$, 
the gauge transformation $\delta_\xi g_{tt}$ of a component $g_{\alpha\beta} t^\alpha t^\beta$ is 
\be
   \delta_{\bm\xi} g_{tt} = \Lie_{\bm\xi} g_{tt} = (\Lie_{\bm\xi} g_{\alpha\beta}) t^\alpha t^\beta 
		\neq \Lie_{\bm\xi} (g_{\alpha\beta} t^\alpha t^\beta).
\ee

We now turn to the problem at hand, the behavior of $U$ and $\Omega$ under gauge transformations. 
These are defined in terms of the four-velocity $u^\alpha$.  To define them and $u^\alpha$ as 
fields, to write a change $\delta u^\alpha$ as a vector at a point using the definition (\ref{eq:deltaT}),  
and to define the Lie derivative $\Lie_\xi u^\alpha$, one must introduce a set of nearby orbits.  
In the problem considered here, one can define an unperturbed vector field $u^\alpha$
by taking $u^{\alpha}(P)$ to be the four-velocity of the circular orbit through $P$.
The perturbed vector field at $P$ is then the four-velocity of the perturbed circular orbit through 
$P$.   
The scalar $U$ of Sec.~\ref{sec4} is constructed locally from $g_{\alpha\beta}, \Omega, t^\alpha$, and 
$\phi^\alpha$: 
\be
  U = {\cal U}[g_{\alpha\beta}(t^\alpha+\Omega\phi^\alpha)(t^\beta+\Omega\phi^\beta)]^{-1/2}
	=:{\cal U}(g_{\alpha\beta},\Omega, t^\alpha,\phi^\alpha).
\ee
Here $t^\alpha$ and $\phi^\alpha$ are reference fields, defined as the vectors ${\bm\partial}_t$ and 
$\bm\partial_\phi$ tangent to the coordinate lines, so that they have the same values for the perturbed 
and unperturbed metric. (Note that they are not Killing fields of the perturbed metric, and cannot 
be defined in this way.)  If $\xi^\alpha$ is helically symmetric with respect to 
$k^\alpha = t^\alpha+\Omega(P)\phi^\alpha$, for a given fixed $P$, then 
\be
 \delta_{\bm\xi} U(P) = \left.\frac d{d\lambda} {\cal U}(g_{\alpha\beta}+\lambda\Lie_\xi g_{\alpha\beta}, \Omega+\lambda\Lie_\xi \Omega, t^\alpha,\phi^\alpha)\right|_{\lambda=0} = \Lie_{\bm\xi} U (P).  
\label{eq:deltaU}\ee  
Similarly, 
\be
 \delta_{\bm\xi} \Omega(P) = \Lie_{\bm\xi} \Omega (P).  
\label{eq:deltaOmega1}\ee 
Finally, defining ${\cal U}_0$ as in Eq.~(\ref{eq:U0}), we have 
\be
  \delta_{\bm\xi} [U-{\cal U}_0(\Omega)] = \left.[\Lie_{\bm\xi} U -\Lie_{\bm\xi} U_0]\right|_{\lambda=0} 
		= \Lie_{\bm\xi} U_0 -\Lie_{\bm\xi} U_0 =0.
\ee
That, is, $U-{\cal U}_0(\Omega)$ is gauge invariant in the usual sense, for helically symmetric gauge vectors.  

 If we denote by $\widehat\delta_{\bm\xi} U$ and $\widehat\delta_{\bm\xi} \Omega$ the changes in 
$U$ and $\Omega$ introduced by Sago et al., Eqs.~(\ref{eq:deltaU}) and (\ref{eq:deltaOmega1})  
are equivalent to  writing
\be 
  \widehat\delta_{\bm\xi} U = 0, \quad  \widehat\delta_{\bm\xi} \Omega=0. 
\label{eq:widehatUO}\ee
Although any scalar constructed locally from physical fields alone has this behavior, 
Eqs.~(\ref{eq:deltaU}) and (\ref{eq:deltaOmega1}) (or, equivalently, (\ref{eq:widehatUO})) are not 
trivial relations, because they involve reference fields that do not change 
under gauge transformations.  

The difference between $\delta_{\bm\xi}$ and $\widehat\delta_{\bm\xi}$ is the change 
associated with a second kind of gauge freedom, mentioned in Appendix A.2 of Paper I, 
namely an infinitesimal change (by a displacement $\xi^\alpha$) in the background geodesic 
to which one compares a geodesic in the perturbed spacetime.  

\bibliography{sf3}
\end{document}